\documentclass[aps,prd,superscriptaddress,nofootinbib,amsmath,amsfonts,preprintnumbers,groupedaddress,10pt,english]{revtex4}
\usepackage{amsmath}
\usepackage{amssymb}
\usepackage{babel}
\usepackage{wrapfig}
\usepackage{cancel}

\usepackage{relsize,exscale}
\makeatletter

\usepackage{array,multirow,graphicx}
\usepackage{dcolumn}
\usepackage{newlfont}
\usepackage{bm}
\usepackage[colorlinks,citecolor=blue,urlcolor=blue,linkcolor=blue]{hyperref}
\usepackage[figtopcap]{subfigure}
\usepackage{color}
\usepackage{verbatim}

\usepackage{scalerel}
\usepackage{tikz}
\usetikzlibrary{svg.path}
\definecolor{orcidlogocol}{HTML}{A6CE39}
\tikzset{
  orcidlogo/.pic={
    \fill[orcidlogocol] svg{M256,128c0,70.7-57.3,128-128,128C57.3,256,0,198.7,0,128C0,57.3,57.3,0,128,0C198.7,0,256,57.3,256,128z};
    \fill[white] svg{M86.3,186.2H70.9V79.1h15.4v48.4V186.2z}
                 svg{M108.9,79.1h41.6c39.6,0,57,28.3,57,53.6c0,27.5-21.5,53.6-56.8,53.6h-41.8V79.1z M124.3,172.4h24.5c34.9,0,42.9-26.5,42.9-39.7c0-21.5-13.7-39.7-43.7-39.7h-23.7V172.4z}
                 svg{M88.7,56.8c0,5.5-4.5,10.1-10.1,10.1c-5.6,0-10.1-4.6-10.1-10.1c0-5.6,4.5-10.1,10.1-10.1C84.2,46.7,88.7,51.3,88.7,56.8z};}}
\newcommand\orcid[1]{\href{https://orcid.org/#1}{\mbox{\scalerel*{
\begin{tikzpicture}[yscale=-1,transform shape]
\pic{orcidlogo};
\end{tikzpicture}
}{|}}}}
\graphicspath{{./}{Figs/}}
\begin{document}
\date{\today}
\title{Properties of compact objects in quadratic non-metricity gravity}

\author{G.~G.~L.~Nashed~\orcid{0000-0001-5544-1119}}

\email{nashed@bue.edu.eg}
\affiliation {Centre for Theoretical Physics, The British University in Egypt, P.O. Box
43, El Sherouk City, Cairo 11837, Egypt}
\affiliation {Center for Space Research, North-West University, Potchefstroom 2520, South Africa}

\author{Kazuharu Bamba~\orcid{0000-0001-9720-8817}}

\email{bamba@sss.fukushima-u.ac.jp}
\affiliation {Faculty of Symbiotic Systems Science, Fukushima University, Fukushima 960-1296, Japan.}

\begin{abstract}
Astrophysical compact objects are studied in the context of quadratic non-metricity gravity. The solutions to the gravitational field equations, which include fluid components, are analyzed to investigate the density and pressure properties of radio pulsars. It is explicitly demonstrated that the theoretically stable models are consistent with astronomical data, due to the geometric features of the quadratic component. Furthermore, it is shown that, in contrast to the compactness limits of black holes in general relativity, the core density can significantly exceed the density at which nuclear saturation occurs, and the surface density can also surpass the value of nuclear saturation. Additionally, it is found that the radial sound speed remains below the conformal upper bound for sound velocity established by perturbative quantum chromodynamics.
\end{abstract}
\maketitle

\section{Introduction}\label{Sec:Introduction}

Understanding the structure and properties of dense celestial objects is a major challenge in general relativity. In 1916, Karl Schwarzschild calculated the solution for the interior of a spherical object with constant density and zero pressure at its surface. Known as the Schwarzschild interior solution, it describes a static, spherically symmetric distribution of matter within a star under general relativity \cite{schwarzschild1916gravitationsfeld}. Although simple and limited to specific types of stars, this solution sparked great interest and led to further studies and extensions \cite{florides1974new,Gron:1985jx,Kohri:2001jx,Gabbanelli:2019txr,Posada:2018agb,Calmet:2020tlj}.

Key advancements came from Tolman \cite{Tolman:1934za,Tolman:1939jz} and Oppenheimer and Volkoff \cite{Oppenheimer:1939ne}, who made significant contributions to relativistic astrophysics. They developed the structural equations for static, spherically symmetric objects in general relativity, including the Tolman-Oppenheimer-Volkoff (TOV) equation, which governs hydrostatic equilibrium in compact stars. Using both numerical and theoretical methods, analysis of the TOV equation determined the maximum mass of neutron stars to be around $3.2M_{\odot}$ \cite{Rhoades:1974fn}. This result was obtained by applying a restrictive equation of state, and it is constrained by the causality condition\footnote{
This condition is given by $P=\rho c^2$, where $P$ and $\rho$ are the pressure and energy density, respectively, and $c$ is the speed of light.}
and the Le Chatelier principle. This consequence also holds if the equation of state (EoS) is uncertain for specific densities.

In comparison, Chandrasekhar \cite{Chandrasekhar:1931ftj} calculated the maximum mass of white dwarfs to be approximately $1.4 M_{\odot}$. Over time, theoretical and observational evidence supported the widely accepted view that neutron star masses cluster around $1.4M_{\odot}$ \cite{Shapiro:1983du}. This value arises from the role of neutron degeneracy pressure, which counteracts gravitational collapse after a white dwarf's collapse. A neutron star with a mass of about $1.4M_{\odot}$ is expected to have a radius between 10 and 15 km and an average density of approximately $6 \times 10^{14} \mathrm{g/cm^3}$.

 Recent advancements have significantly reshaped traditional interpretations of neutron star masses, thanks to more precise measurements \cite{Horvath:2021ang}. Observations, combined with the discovery of gravitational waves, have demonstrated that neutron star masses vary much more than previously thought, surpassing the limits suggested by the Chandrasekhar limit. In Ref.~\cite{Margalit:2017dij}, scientists determined that the maximum mass of a neutron star is about $2.17 M_\odot$. This limit was derived by using the data of electromagnetic and gravitational waves associated with the binary neutron star merger GW170817. The constraints from these observations are stricter and less model-dependent, compared to other estimates.

Shibata et al. \cite{Shibata:2017xdx} analyzed this further, assuming a rigid equation of state for neutron matter. They concluded that neutron stars must have masses exceeding $2 M_\odot$ to remain stable after mergers such as GW170817, in which the combined mass exceeded $2.73M_\odot$. If no relativistic optical counterpart is observed, the maximum neutron star mass is estimated to lie between $2.15M_\odot$ and $2.25M_\odot$. Similar results were obtained by Ruiz and Rezzolla \cite{Ruiz:2017due,Rezzolla:2017aly}, who provided comparable upper mass limits. Pulsar mass measurements using techniques like Shapiro delay have yielded precise estimates, including $1.928 \pm 0.017 M_\odot$ for PSR J1614-2230 \cite{Fonseca:2016tux} and $2.14^{+0.10}_{-0.09} M_\odot$ for the millisecond pulsar MSP J0740+6620 \cite{NANOGrav:2019jur}.

Notably, the GW190425 event provided intriguing insights into neutron star masses and suggested a total mass of approximately $3.4 M_{\odot}$ for a binary neutron star system \cite{LIGOScientific:2020aai}. Similarly, data from GW190408 indicated the mass of a neutron star in the range of $2.5$-$2.65 M_{\odot}$ merging with a significantly larger black hole of about $26 M_{\odot}$ \cite{LIGOScientific:2020zkf}. These findings offer new perspectives on neutron star masses and their distribution, challenging the conventional assumption that a standard value of a mass is around $1.4M_{\odot}$.


One possible explanation for the higher mass of compact objects is the inclusion of exotic components. Harko (2009) \cite{Harko:2009ysn} demonstrated that a rotating quark star in the Color-Flavor-Locked (CFL) phase could have a mass range of $3.8$-$6M_{\odot}$, making it resemble a stellar-mass black hole. However, detecting such stars is challenging due to their low luminosity. On the other hand, the so-called Bose-Einstein Condensate (BEC) stars consist of the condensation of superfluids originated from particles with their double masses of a neutron. These particles form Cooper pairs and exhibit the scattering lengths of $10$-$20$ fm, resulting in maximum masses of roughly $2M_{\odot}$. If kaons form the condensate, the neutron star's mass could range between $2.4$ and $2.6M_{\odot}$ \cite{Chavanis:2011cz}. For further discussion on the impact of exotic components like quarks and kaon condensates on neutron star interiors, see Ref.~\cite{Glendenning:1997wn}. An important approach to understanding higher pulsar masses involves exploring modified gravitational theories. These theories have emerged to address phenomena such as the accelerated expansion of the universe \cite{Sotiriou:2008rp,Capozziello:2011et,Nojiri:2010wj,Joyce:2014kja,Nojiri:2017ncd,Frusciante:2019xia}.

Extending general relativity through modified gravity theories offers valuable insights into the structure of dense astronomical objects. In these extended theories, key stellar characteristic including the mass-radius relation, the maximum value of a mass, and moments of inertia are significantly different from general relativity's predictions \cite{Olmo:2019flu}. In particular, even for non-relativistic stars like dwarfs, modified theories of gravitation strongly influence their internal composition due to the changes of gravitational interactions. Such differences arise from modifications to the Poisson equation for gravitational potential, affecting a star's mass, radius, central density, and luminosity \cite{Olmo:2019flu}. For example, such amendments impact the minimum mass required for sustained hydrogen burning in brown dwarfs and the Chandrasekhar mass limit for white dwarfs. Observations of neutron star masses, typically around $2M_{\odot}$, often contradict certain equations of state predictions for compact matter. As a result, many EOS models, particularly those involving hyperons, have been ruled out \cite{Olmo:2019flu}. By incorporating adjustments to the equations of the hydrostatic equilibrium in stellar models of stars, alternative gravity theories could help address or resolve these discrepancies \cite{Olmo:2019flu}.

Various modified gravity theories were studied to explain the properties of compact astrophysical objects including neutron stars and the distribution of the masses. For instance, $f({\mathcal R},{\mathcal G})$, $f({\mathcal G})$, $f({\mathcal R},L_m)$, $f({\mathcal R},{\mathcal T})$, hybrid-metric Palatini gravity, and theories that extend the standard Hilbert-Einstein Lagrangian by incorporating a three-form field $A_{\alpha \beta \gamma}$ \cite{Olmo:2019flu}. In these modified gravity frameworks, the mass of the static and spherically symmetric objects can be calculated, although this becomes especially challenging in general relativity when rapid rotation effects are considered. In Ref.~\cite{Ong:2013qja}, it has been shown that the general formulation of $f(T)$ gravity may encounter issues with superluminal propagating modes. Additionally, in the context of flat spatial cosmology, it has been shown that symmetric teleparallel gravity can exhibit intrinsic ghost modes \cite{Gomes:2023tur}, though these ghost modes have not yet been identified in other symmetry settings.

The $f(\mathcal{Q})$ gravity theory has emerged as an intriguing class of extended theories of gravity, offering new insights into both astrophysical and cosmological phenomena \cite{Heisenberg:2023lru,BeltranJimenez:2018vdo,Gomes:2023tur,BeltranJimenez:2017tkd,DAmbrosio:2021zpm,Casalino:2020kdr, Zhao:2021zab,Lazkoz:2019sjl,Mandal:2020lyq,Capozziello:2022tvv,Capozziello:2022wgl}. Through recent studies, its similarities as well as differences have been investigated in comparison with $f(\mathcal{R})$ and $f(\mathcal{T})$ gravity, particularly in terms of their degrees of freedom and symmetry-breaking characteristics, providing new approaches to understanding late-time cosmic acceleration \cite{Hu:2022anq}. Modified non-metricity theory plays a crucial role in astrophysics, offering a fresh perspective on the universe's fundamental structure and the behavior of gravity in extreme environments. This framework enables the study of compact astrophysical objects, the mass distribution, and the equation of state for high-density matter. Additionally, non-metricity gravity provides a valuable framework for exploring the late-time expansion of the universe, a key topic in cosmology. Notably, the application of the Arnowitt-Deser-Misner (ADM) formulation to $f(\mathcal{Q})$ gravity demonstrates the ability to eliminate ghost scalar modes, enhancing the theory's stability \cite{Hu:2023gui}. Furthermore, a new model has been developed to incorporate the inflationary epoch within the $f(\mathcal{Q})$ framework, showcasing its versatility in addressing cosmological challenges \cite{Nojiri:2024zab}. The ability of $f(\mathcal{Q})$ gravity to tackle complex gravitational interactions and solve persistent astrophysical mysteries makes it a vital tool for researchers. Its potential to bridge theoretical models with observational data and open new pathways for exploring the universe underscores its importance in modern astrophysical research.

In this study, we investigate the geometric effects of higher-order terms in non-metricity gravity on the properties of astrophysical compact objects. Key aspects under consideration include the mass-radius relation, energy conditions, stability criteria, and the induced equation of state. Comparisons with compact objects in $f(R)$ gravity \cite{Nashed:2024pbc} and higher-order Gauss-Bonnet gravity \cite{Nashed:2024dno} reveal the following insights.
(i) Compactness Constraint: In quadratic non-metricity gravity, the maximum compactness of compact objects does not exceed the Buchdahl limit, which represents the black hole threshold. This result aligns with higher-order Gauss-Bonnet gravity but contrasts with $f(R)$ gravity, in which the compactness can surpass this limit.
(ii) Radial Sound Speed Limit: A novel feature of quadratic non-metricity gravity is that the radial sound speed remains below the conformal upper limit derived from perturbative quantum chromodynamics, as established in Ref.~\cite{Bedaque:2014sqa}.
These findings highlight the unique astrophysical implications of quadratic non-metricity gravity and its role in providing a constrained and stable framework for describing compact objects.

The organization of this paper is structured as follows.
In Sec.~\ref{sec22}, we explore the non-metricity gravity in which the Lagrangian density is represented as $f(\mathcal{Q})$ and obtain the gravitational field equations for a spacetime with its spherical symmetry under the distribution of an anisotropic matter. In Sec.~\ref{Sec:Model}, we apply the quadratic non-metricity gravity to a stellar model, incorporating the junction conditions and matching with the Schwarzschild exterior solution. In Sec.~\ref{Sec:Stability}, we utilize the mass and radius constraints from astrophysical observations, particularly those related to the pulsar ${\mathcal J0740+6620}$, to obtain the observational constrains on the quadratic non-metricity gravity. Additionally, we assess the model's validity through various stability criteria in both the matter and geometric sectors. In Sec.~\ref{Sec:EoS_MR}, we derive the equations of state (EoS) governing the matter component in the stellar model. We finally conclude this study and summarize the key findings and implications of the results in Sec.~\ref{Sec:Conclusion}.
Throughout this paper, we use the notation that $G$ is the gravitational constant and $c$ is the speed of light.

\section{The astrophysical formulae in $f(\mathcal{Q})$ gravity}\label{sec22}

In this section, the astrophysical formulae are briefly described in $f(\mathcal{Q})$ gravity.

\subsection{Teleparallel $f\left(  \mathcal{Q}\right)$ gravity}\label{sec2}

We consider a four-dimensional space with a metric tensor $g_{\mu\nu}$ and a covariant derivative $\nabla_{\mu}$ using the general connection $\Gamma_{\mu\nu}^{\kappa}$. The curvature tensor, denoted as $R_{\lambda\mu\nu}^{\kappa}$, is defined as follows
\begin{equation} \label{curvaturedef}
{ R_{\lambda\mu\nu}^{\kappa}\left(  \Gamma_{\mu\nu}^{\lambda}\right)
=\frac{\partial\Gamma_{\lambda\nu}^{\kappa}}{\partial x^{\mu}}%
-\frac{\partial\Gamma_{\lambda\mu}^{\kappa}}{\partial x^{\nu}}%
+\Gamma_{\lambda\nu}^{\sigma}\Gamma_{\mu\sigma}^{\kappa}-\Gamma
_{\lambda\mu}^{\sigma}\Gamma_{\mu\sigma}^{\kappa}}\,.
\end{equation}
Here, the torsion tensor, $T_{\mu\nu}^{\lambda}$ is expressed as
\begin{equation} \label{torsiondef}
{ T_{\mu\nu}^{\lambda}\left(  \Gamma_{\mu\nu}^{\lambda}\right)  =\Gamma
_{\mu\nu}^{\lambda}-\Gamma_{\nu\mu}^{\lambda}}\,,
\end{equation}
and the definition of a non-metricity tensor $\mathcal{Q}_{\lambda\mu\nu}$ is
given by
\begin{equation} \label{nonmdef}
{ \mathcal{Q}_{\lambda\mu\nu}\left(  \Gamma_{\mu\nu}^{\lambda}\right)  = \nabla_{\lambda} g_{\mu\nu}=\frac{\partial
g_{\mu\nu}}{\partial x^{\lambda}}-\Gamma_{\lambda\mu}^{\sigma}g_{\sigma\nu
}-\Gamma_{\lambda\nu}^{\sigma}g_{\mu\sigma}}\,.
\end{equation}

The Levi-Civita connection in general relativity, written as $\hat{\Gamma}_{\mu\nu}^{\kappa}$, is defined as
\begin{equation} \label{LCcon}
  \hat{\Gamma}_{\mu\nu}^{\kappa} = \frac{1}{2} g^{\kappa\lambda} \left( \frac{\partial g_{\lambda\mu}}{\partial x^\nu} + \frac{\partial g_{\lambda\nu}}{\partial x^\mu} - \frac{\partial g_{\mu\nu}}{\partial x^\lambda} \right).
\end{equation}
If we use $\hat{\Gamma}_{\mu\nu}^{\kappa}$ instead of $\Gamma_{\mu\nu}^{\kappa}$ in Eq.~\eqref{curvaturedef},
we get ${\hat{R}_{\lambda\mu\nu}}^{\kappa}= {R_{\lambda\mu\nu}}^{\kappa}\left( \hat{\Gamma}_{\mu\nu}^{\lambda}\right)$.
This can be realized for the case that $\mathcal{Q}_{\lambda\mu\nu}\left(\hat{\Gamma}_{\mu\nu}^{\lambda}\right) = 0$ is set to be zero.
The Ricci scalar is considered to be
the fundamental component, and hence it is defined as
$\hat{R}=g^{\mu\nu}\hat{R}_{\mu\nu}$,
where $\hat{R}_{\mu\nu}=g^{\kappa\nu}R_{\kappa\mu\nu\lambda}$.

For the Weitzenbock connection in teleparallelism, $\tilde{\Gamma}_{\mu\nu}^{\kappa}$, the curvature tensor is vanishing identically, i.e.,  $R_{\lambda\mu\nu}^{\kappa}\left( \tilde{\Gamma}_{\mu\nu}^{\lambda }\right)=0$
and $\mathcal{Q}_{\lambda\mu\nu}\left(\tilde{\Gamma}_{\mu\nu}^{\lambda }\right) =0$.
Therefore, Eq.~\eqref{torsiondef} gives the nonzero torsion
$T_{\mu\nu}^{\lambda}\left(\tilde{\Gamma}_{\mu\nu}^{\lambda}\right)$.
The Lagrangian of teleparallel equivalent of general relativity consists of the torsion scalar,
$T={S}_\kappa^{\mu\nu}T^\kappa_{\mu\nu}$, where  $S_\beta^{\mu\nu}=\frac{1}%
{2}(K^{\mu\nu\,}_{ \, \,\,\,\beta}+\delta_{\beta}^{\mu}T^{\theta\nu}_{\,\,\,\, \theta}
-\delta_{\beta}^{\nu}T^{\theta\mu}_{\,\,\,\,\theta})$ and $K_{\beta}^{\mu\nu
}=-\frac{1}{2}(T^{\mu\nu}_{\,\,\, \beta}-T^{\nu\mu}_{\,\,\,\,\beta}-T_{\beta}^{.\mu\nu
})$ \cite{Ferraro:2006jd}.

In the symmetric teleparallel gravity,
$R_{\lambda\mu\nu}^{\kappa}\left( \Gamma_{\mu\nu}^{\lambda }\right) = 0$ represents a flat geometry without torsion,
and $T_{\mu\nu}^{\lambda}\left( \Gamma_{\mu\nu }^{\lambda}\right) =0$, whereas \eqref{nonmdef} defines the tensor of non-metricity. The scalar of non-metricity  provides the fundamental Lagrangian density of symmetric teleparallel gravity \cite{Nester:1998mp}
\begin{equation}
\mathcal{Q}=\mathcal{Q}_{\lambda\mu\nu}P^{\lambda\mu\nu}\,.\label{defQ}%
\end{equation}
Here, $P_{\mu\nu}^{\lambda}$ represents the components of the conjugate tensor, and it is given by
\begin{equation}
P_{\mu\nu}^{\lambda}=-\frac{1}{4}\mathcal{Q}_{\mu\nu}^{\lambda}+\frac{1}{2}%
\mathcal{Q}_{(\mu\phantom{\lambda}\, \nu)}^{\phantom{(\mu}\lambda \phantom{\nu)}}+\frac
{1}{4}\left(  \mathcal{Q}^{\lambda}-\bar{\mathcal{Q}}^{\lambda}\right)  g_{\mu\nu}-\frac{1}%
{4}\delta_{\;(\mu}^{\lambda}\mathcal{Q}_{\nu)},\label{defP}%
\end{equation}
where the Kronecker delta is represented by $\delta_{\;\nu}^{\mu}$, and the contracted tensors are introduced as $\mathcal{Q}_{\mu}=\mathcal{Q}_{\mu\nu}^{\phantom{\mu\nu}\nu}$ and
$\bar{\mathcal{Q}}_{\mu}=\mathcal{Q}_{\phantom{\nu} \mu\nu}%
^{\nu\phantom{\mu}\phantom{\mu}}$.
Beginning with the teleparallel condition $R=0$, which describes a flat geometry associated with a purely inertial connection, we can apply a linear group \textit{gauge} transformation ${\textrm GL }(4, {\cal \mathbb{R}})$ parameterized by $\Lambda_{\phantom{\alpha}\mu}^{\alpha}$ \cite{BeltranJimenez:2019esp,BeltranJimenez:2019tme},
\begin{equation}
\Gamma_{\phantom{\alpha}\mu\nu}^{\alpha}=\left(\Lambda^{-1}\right)_{\phantom{\alpha}\beta}^{\alpha}\partial_{[\mu}\Lambda_{\phantom{\alpha}\nu]}^{\beta}.
\end{equation}
Thus, the most general possible connection can be expressed using the general element of ${\textrm GL}(4, {\cal \mathbb{R}})$, which is parameterized by the transformation
$\Lambda_{\phantom{\alpha}\mu}^{\alpha}=\partial_{\mu}\xi^{\alpha}$, where $\xi^{\alpha}$ is an arbitrary vector field.,
\begin{equation}
\Gamma_{\phantom{\alpha}\mu\nu}^{\alpha}=\frac{\partial x^{\alpha}}{\partial\xi^{\rho}}\partial_{\mu}\partial_{\nu}\xi^{\rho}. \label{coinc}
\end{equation}
This result demonstrates that the connection can be eliminated through a coordinate transformation. The transformation that removes the connection in Eq. \eqref{coinc} is referred to as \emph{{\color{blue} the gauge coincident}} \cite{BeltranJimenez:2017tkd}.
\par
As a result, in the coincident gauge, the non-metricity tensor defined by Eq. \eqref{nonmdef} simplifies to:
\begin{equation}
 Q_{\beta\mu\nu}\equiv \partial_{\beta}g_{\mu\nu}. \label{tns_nmetric2}
\end{equation}
In this manuscript, we adopt the coincident gauge to compute our solutions.

\subsection{Equations of motions}
The gravitational action integral in symmetric teleparallel $f\left( \mathcal{Q}\right) $ gravity is expressed as
\begin{equation}
S=\frac{1}{2\kappa}\int d^{4}x\sqrt{-g}f(\mathcal{Q})+\int d^{4}x\sqrt{-g}\mathcal{L}_{M}\,.
\label{action}%
\end{equation}
Here, $g=\mathrm{det}(g_{\mu\nu})$ refers to the determinant of the spacetime metric, while $\mathcal{L}_{M}$ defines the Lagrangian that characterizes the matter content. Moreover, we denote $\kappa=8\pi G/c^4$.
The gravitational field equation, derived by the variation of the metric,
reads \cite{Xu:2019sbp}
\begin{equation} \label{fieldm}
f^{\prime}(\mathcal{Q})G_{\mu\nu}+\frac{1}{2}g_{\mu\nu}\left[  f^{\prime}%
(\mathcal{Q})\mathcal{Q}-f(\mathcal{Q})\right]  +2f^{\prime\prime}(\mathcal{Q})\left(  \nabla_{\lambda}\mathcal{Q}\right)
P_{\;\mu\nu}^{\lambda}=\kappa T_{\mu\nu}\,.
\end{equation}
with $ \nabla_{\lambda}\mathcal{Q}\equiv \partial_\lambda \mathcal{Q}$ {\color{blue} and $G_{\mu\nu}$ is the Einstein tensor defined as $G_{\mu\nu}=R_{\mu\nu}-1/2r_{\mu\nu}R$.}
Here, $f^{\prime}(\mathcal{Q})$ means the derivative of $f$ in terms of $\mathcal{Q}$ and the primes denote derivatives taken with respect to the respective arguments. Furthermore, $T_{\mu\nu}$ expresses the energy-momentum tensor associated with the matter source.
We note that when the function $f\left(  \mathcal{Q}\right)$ is linear or constant, the field equations are simplified as the ones which have an additional term corresponding to the cosmological constant in general relativity. Therefore, to investigate theories and dynamics that extend beyond general relativity, we will limit our analysis to the case that $\mathcal{Q}$ is not a constant and thus $f(\mathcal{Q})$ has an appropriate nonlinear form.

The evolution of the connection is governed by the following equation
\begin{equation}
{\mathcal \nabla_{\mu}\nabla_{\nu}\left(  \sqrt{-g}f_\mathcal{Q} P_{\phantom{\mu\nu}\sigma}^{\mu\nu}\right) } =0\,,
\label{feq2}%
\end{equation}
{\color{blue} where $f_\mathcal{Q}=\frac{df(\mathcal{Q})}{d\mathcal{Q}}$.} 
The equation of motion for a matter source with a minimal coupling to the metric is given by $\nabla_{\mu} T_{\phantom{\mu}\nu}^{\mu}=0$, where
where $\nabla_{\mu}$ signifies the covariant derivative concerning the Levi-Civita connection. The equation of motion for a matter source with a minimal coupling to the metric is given by $T_{\phantom{\mu}\nu;\mu}^{\mu}=0$, where the semicolon \textquotedblleft$;$\textquotedblright\ signifies the covariant derivative concerning the Levi-Civita connection.

In this study, we concentrate on the case that the spacetime consists of the anisotropic matters. Hence, the total energy-momentum tensor ($T^\mu{}_\nu$) can be given by
\begin{eqnarray}
T^\mu{}_\nu=\epsilon\,u^\mu\,u_\nu-\mathcal{P}\,K^\mu{}_\nu+\Pi^\mu{}_\nu, \label{eq17}
\end{eqnarray}
where
\begin{eqnarray}
&&\hspace{-0.5cm} \mathcal{P}=\frac{P_r+2P_\perp}{3},\qquad \qquad \Pi^\mu{}_\nu=\Pi \big(\zeta^\mu \zeta_\nu+\frac{1}{3} K^\mu{}_\nu\big),\nonumber\\&& \text{with}~~ \Pi=P_r-P_\perp,  \qquad \qquad K^\mu{}_\nu=\delta^\mu{}_\nu-u^\mu u_\nu. \label{eq18}
\end{eqnarray}
Here, the four-velocity vector of a fluid $u^\mu$ and the unit space like vector $\zeta^\mu$ (with $\{\mu=0,1,2,3\}$) read
\begin{eqnarray}
u^\mu=(\sqrt{g^{tt}},~0,~0,~0)~~\text{and}~~\zeta^\mu=(0,~\sqrt{g^{rr}},~0,~0),~~ \label{eq19}
\end{eqnarray}
respectively, so that $\zeta^\mu u_\mu=0$ and $\zeta^\mu\zeta_\mu=-1$. Moreover, $\epsilon$ shows the total energy density. On the other hand, $P_r$ and $P_{\perp}$ mean the total radial and tangential pressure for the gravitationally decoupled system, respectively.

For the spherically symmetric line element, in the decoupled system gravitationally, the components of the energy-momentum tensor are expressed as
\begin{eqnarray}
&& T^0_0=\epsilon,~~~T^1_1=-P_r,~~~~T^2_2=T^3_3=-P_\perp . \label{eq20}
\end{eqnarray}

In the next section, we delve into examining specific instances of spacetimes. Our goal is to explore how well the $f(\mathcal{Q})$ theory aligns with the interior spherically symmetric solution.

\section{The solutions of anisotropic spacetimes in quadratic non-metricity gravity}\label{Sec:Model}
In this section, to find a solution for an anisotropic spacetime,
we investigate the following quadratic Lagrangian density
\begin{equation}\label{VM}
f(\mathcal{Q}) =\mathcal{Q}+\xi \mathcal{Q}^2\,,
\end{equation}
where $\xi$ represents a dimensional quantity with the unit of ${\textit length}^2$.

\subsection{Metric to describe an anisotropic spacetime}\label{Sec:IIIA}
We take the ansatz of the metric, represented as
\begin{align}\label{RG}
   & ds^2 = g_{\mu \nu} dx^\mu dx^\nu=-e^{a(r)}c^2dt^2 + e^{b(r)}dr^2 + r^2d\Omega^2\,, \qquad  \nonumber\\
   & \mbox{where} \quad d\Omega^2= d\theta^2 + \sin^2\theta d\phi^2 \,,
\end{align}
with $x^\mu$ the components of the vector in four positions, including radial distance ($r$), time ($t$), and angular coordinates ($\theta$ and $\phi$).  The metric functions, represented as $a$ and $b$, are two functions that depend solely on the radial coordinate $r$. Consequently, we can calculate $\sqrt{-g}=r^2 \sin \theta, \, e^{(a + b)/2}$ as well as the four-velocity vector $v^\mu=(ce^{-a/2}, 0, 0, 0)$.

Through the utilization of Eq.~(\ref{RG}),
we derive the non-metricity scalar in the following manner
\begin{align}\label{RaG}
\mathcal{Q}={\frac { \left( {e^b}-1 \right) {e^{-b
  }} \left( a' +b'  \right) }{r}}\,.
\end{align}
Here and in the following, we use the notation such as $'\equiv d/dr$ and $''\equiv d^2/dr^2$.
By using the expression of the non-metricity in Eq.~(\ref{RaG}),
from the field equations \eqref{fieldm} with (\ref{VM}), we achieve
the representations of $\epsilon$, $P_r$ and $P_\perp$,
which are shown in Appendix~\ref{Sec:Appendix New A.}.
It is clearly seen that the density and pressures of matter undergo changes and approach the solution in general relativity when $\xi_1$ is taken as 0 (as explained in Refs.~\cite{Nashed:2020kjh, Roupas:2020mvs}).

\subsection{The metric potentials of anisotropic spacetimes}\label{Sec:KB}

Due to the fact that Eqs.~(\ref{Rho1}), (\ref{pr1}), and (\ref{pt1}) create a complex system of three nonlinear differential equations with five variables - $a$, $b$, $\rho$, $P_r$, and $P_\perp$ - further restrictions are needed to solve the system. One way to enforce these restrictions is by using particular equations of state (EoS) as an approach.  Nevertheless, this approach may not be very beneficial as the systems mentioned above, which are Eqs.~(\ref{Rho1}), (\ref{pr1}), and (\ref{pt1}), consist of non-linear differential equations.
The best strategy is to recommend precise forms for the ansatz of $a$ and $b$. This investigation will make use of the Krori-Barua (KB) metric potentials described in Ref.~\cite{Krori1975ASS}, which are formulated as\footnote{
It is well know that the KB metric is studied before and confront with observations and give a consistent results with observations \cite{Nashed:2024pbc,Nashed:2023uvk,Nashed:2023lqj}
}
\begin{equation}\label{eq:KB}
   {  a(r)=s_0 (r/L_s)^2+s_1,\,  \qquad b(r)=s_2 (r/L_s)^2}\,.
\end{equation}
In this context, $L_s$ refers to the star's radius, and by fulfilling matching conditions, we can determine the values for the dimensionless parameters ${s_0, s_1, s_2}$. These equations ensure a steady and well-behaved solution spanning the star completely. The ansatz of the metric potential given by Krori and Barua has found application in diverse altered theories as well as in  general relativity. However, in the present study, we use the constraints set by recent NICER observations concerning the pulsar ${\mathcal J0740+6620}$, its mass and radius to fix the coefficient of the quadratic form of the non-metricity, represented as $\xi_1$. To unify the unit adjustments, it is practical to employ the dimensionless parameter in the manner described below
\begin{equation}\label{const}
\xi=\xi_1L_s{}^2.
\end{equation}
By instituting Eqs.~\eqref{eq:KB} and \eqref{const} into Eqs.~(\ref{Rho1}), (\ref{pr1}) and (\ref{pt1}), we obtain exact expressions of the two components of pressures $P_r$ and $P_\perp$ in addition to the energy density $\epsilon$.
We describe these representations in Appendix~\ref{Sec:Appendix New B.}.

The anisotropy is expressed as $F_a=\frac{2\Delta}{r}$, which arises from the difference of pressures, expressed by the anisotropic parameter $\Delta={P_\perp}-{P_r}$. It is crucial to note that $\Delta$ vanishes in the core. For the strong anisotropy as $0<r\leq L_s$, $P_\perp$ must be greater than $P_r$ in all region  of the star. Conversely, when $\Delta<0$, which represents mild anisotropy, it is crucial that $P_r$ should be greater than $P_\perp$ throughout the star.

\subsection{Boundary conditions to connect the interior metric to the Schwarzschild exterior one}\label{Sec:Match}

The star's geometry, whether shaped by internal forces or external factors, does not affect the boundary metric. Consequently, the metric tensor components will remain continuous across the boundary surface, no matter which coordinate system is applied. In the context of general relativity, the Schwarzschild solutions have been instrumental in shaping the selection of appropriate matching conditions when analyzing compact stellar objects. In modified gravity theories, the exterior solution of a star may differ from the Schwarzschild solution when the TOV equations are altered and pressure and energy density vanish. However, by choosing an appropriate \( f(\mathcal{Q}) \) gravity model, the solutions of the altered TOV equations include non-zero energy and pressure could still align with the Schwarzschild solution. This might be the cause why Birkhoff's theorem is not applicable in the realm of amended gravitational theories. A careful examination of this issue in the frame of $f({\mathcal{Q}})$ theory would be highly interesting. The solution of the Schwarzschild spacetime in $f({\mathcal{Q}})$ gravity has been explored and a range of fascinating findings was found \cite{Cooney:2009rr, Goswami:2014lxa,Ganguly:2013taa,Astashenok:2014pua}\footnote{ Indeed, during the investigation of the junction conditions in $f(\mathcal{Q})$ gravity, Lin et al. have shown that the external background solutions for any regular form of $f({\mathcal{Q}})$ coincide with the corresponding solutions in general relativity~\cite{Lin:2021uqa}.}.


We can study the matching conditions by seamlessly merging the entire  metric at $r = L_s$ with exterior solution  as represented by\footnote{Due to the fact that the field equation reproduced by the action in Eq.~(\ref{action}), which does not include a cosmological constant, we will use the exterior solution given by the Schwarzschild solution \cite{Lin:2021uqa}.}
\begin{equation}
 {ds^2=-\left(1-\frac{2GM}{c^2r}\right) c^2 dt^2+\frac{dr^2}{\left(1-\frac{2GM}{c^2 r}\right)}+r^2 (d\theta^2+\sin^2 \theta d\phi^2)}\,,
\end{equation}
with $M$ signifies the mass of pulsar.
Applying the entire spacetime in Eq.~\eqref{eq:KB},
the matching condition is implemented at the boundary of the pulsar.
Consequently, we find
\begin{align}\label{eq:bo}
 &{\mathrm a(r={L_s})=\ln(1-C)\,, \qquad b(r={L_s})=-\ln(1-C)\,, \qquad \text{and}} \, \nonumber\\
 &{\mathrm{{P}_r}(r={L_s})=0}\,.
\end{align}
The compactness is represented by the following form
\begin{align}\label{comp11}
 {  C=\frac{2GM}{c^2 {L_s}}}\,.
\end{align}
Through the use of radial pressure presented in Appendix~\ref{Sec:Appendix New B.} and the KB given by Eq.~\eqref{eq:KB}, together with the mentioned boundary conditions, we can express the model parameters $[s_0, s_1, s_2]$ expressed in relation to $\xi_1$ and the parameters signifying compactness. Thus, the parameter of $f({\mathcal{Q}})$, that takes the quadratic form, essentially includes $[\xi_1, C]$.  Considering that astrophysical observations impose constraints on the  radius and mass of stars, influencing their compactness, the next task is to define constraints on  $\xi_1$.

\section{Mass-radius relation and the stability conditions of astrophysical pulsars}\label{Sec:Stability}

We utilize the observed limits on $M$ and $R$ of star ${\mathcal J0740+6620}$ to evaluate the estimated numerical value of $\xi_1$. We also assess wither the model under investigation is stable or not considering various physical constraints. Firstly, it is essential to discuss our choice of ${\mathcal PSR J0740+6620 }$ in order to constrain quadratic $f({\mathcal{Q}})$ gravity specifically. We can use the Shapiro time delay to measure the sellar mass, which is found to be $M= 2.08 \pm 0.07 M_\odot$ \cite{NANOGrav:2019jur,Fonseca:2021wxt}.It is crucial to emphasize that the star's mass approaches the maximum limit typically expected for a neutron star, where significant changes in terms of gravitation can occur.
In addition, the accuracy of weight measurement is very high and does not depend on the tilt angle, mainly because this celestial body belongs to a pair of stars. Moreover, gathering information from the X-ray multi-mirror (XMM) Newton dataset increases the accuracy of measurements of the radius of a pulsar, especially in scenarios where the NICER count rate is low $L= 13.7_{-1.5}^{+2.6}$ km \cite{Miller:2021qha}. Another independent examination provides a value of
$L=12.39_{-0.98}^{+1.30}$ km \cite{Riley:2021pdl}. Interestingly, by using a nonparametric form of the EoS that addresses NICER+XMM and employing a Gaussian process, the radius and mass of ${\mathcal PSR J0740+6620 }$ are found to be
$L=12.34^{+1.89}_{-1.67}$ km and
$M=2.07 \pm 0.11 M_\odot$ \cite{Legred:2021hdx}. This new measurement aligns with the findings in Ref.~\cite{Landry:2020vaw} within a 1$\sigma$ level. In this situation, the pulsar ${\mathcal PSR J0740+6620 }$ is an excellent testing ground for narrowing down the parameter space of $f({\mathcal{Q}})$ theory, through the use of  the quadratic form, within the framework of  KB model. This primarily involves the parameters {$\xi_1$, $s_0$, $s_1$, $s_2$}.

\subsection{Astrophysical effects of metric components}\label{Sec:geom}
\begin{figure}
\subfigure[~Identification solutions]{\label{fig:Junction}\includegraphics[scale=0.38]{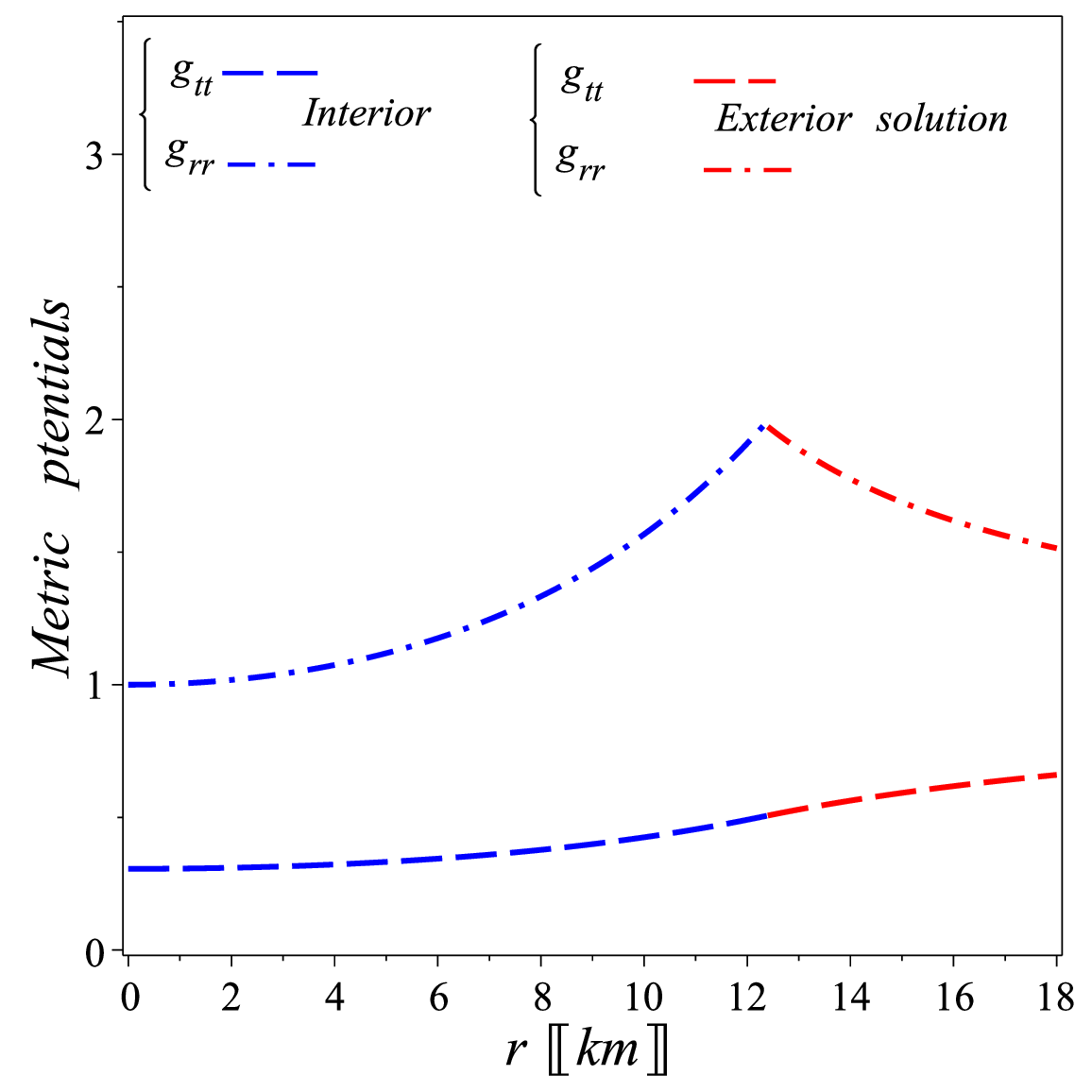}}\hspace{1cm}
\subfigure[~Redshift]{\label{fig:redshift}\includegraphics[scale=0.38]{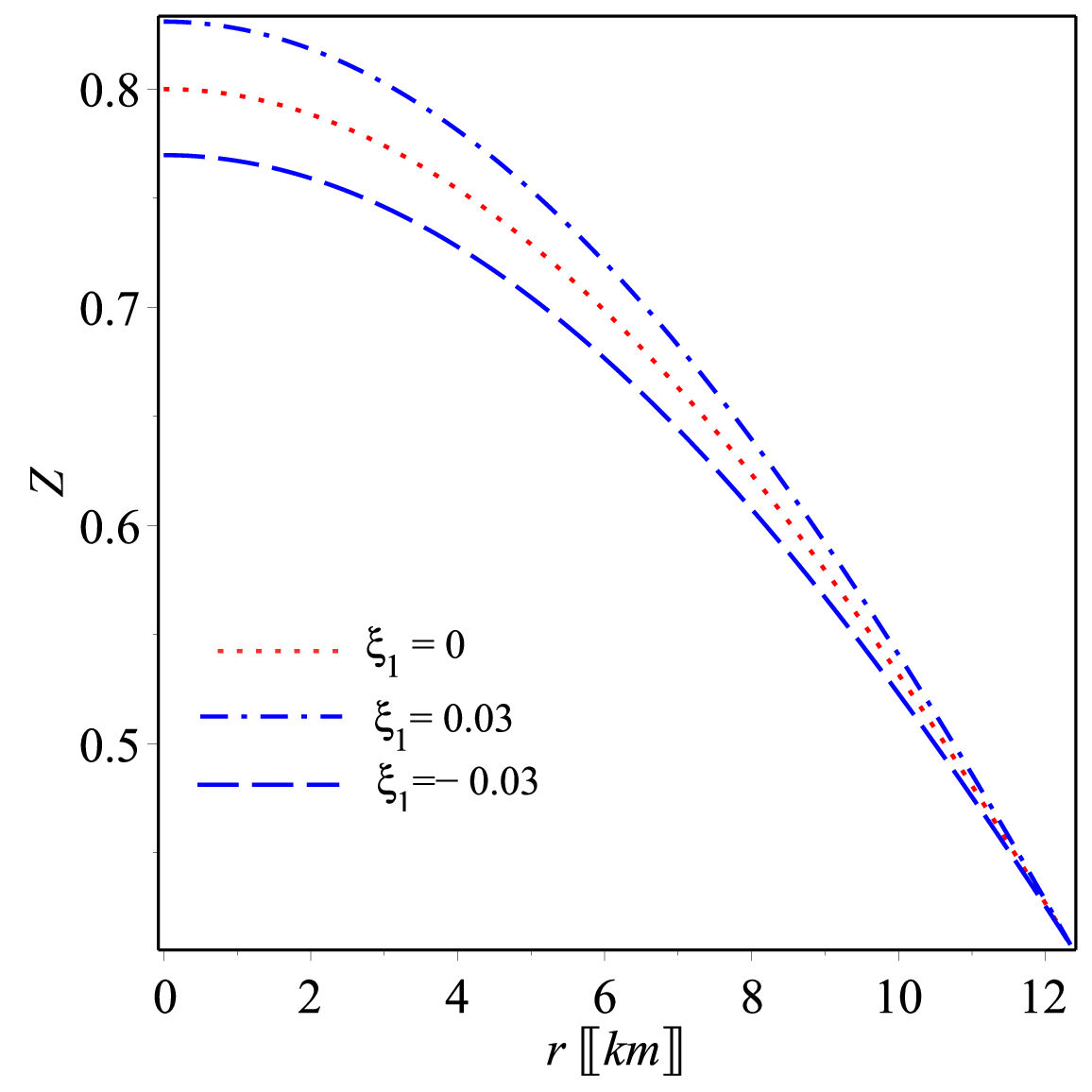}}
\caption{The visual representation of pulsar ${\mathcal J0740+6620}$ in \subref{fig:Junction} displays $g_{tt}$ and $g_{rr}$ inside the star, calculated using KB form, and the outside the star, obtained from the external vacuum solution in the Schwarzschild spacetime. It is seen that the potentials of the metric remain finite values within the interior of pulsars and have a smooth transition to the exterior region; \subref{fig:redshift} shows the pattern of the redshift in the case where $\xi_1$ is either 0 or close to $\pm 0.0004$, then the redshift function in Eq.~\eqref{eq:redshift} shows a peak redshift of approximately 0.8 at the center, decreasing to around 0.4 at the edge of the pulsar in all cases.}
\label{Fig:Matching}
\end{figure}

Highlighting the components of the metric is essential, as both  ${\textit 1/g^{tt}}$ and ${\textit 1/g^{rr}}$ must not show any singularities within the pulsar's interior.  The selected KB equation \eqref{eq:KB} ensures the smoothness of the metric potentials in the central area, i.e.,  ${ \textit 1/g^{tt} (r\to0)=e^{a}}\neq 0$ and ${ 1/g^{rr}(r\to 0)=1}$.
In Fig.~\ref{Fig:Matching} \subref{fig:Junction}, it is illustrated how ${\textit 1/g^{tt}}$ and ${\textit 1/g^{rr}}$  behave at any  position in the star. Moreover, we investigate  the consistency between the internal  solution given by KB and the vacuum  Schwarzschild one at the surface of  ${\mathcal J0740+6620}$, precisely when $L=12.34$ km, utilizing the  value of the parameter that characterizes the model as displayed in Fig.~\ref{Fig:Matching}.

Furthermore, we articulate the function that represents gravitational redshift in relation to the KB model within the framework of $f({\mathcal{Q}})$ in Eq.~\eqref{VM} as follows
\begin{equation}\label{eq:redshift}
  {\mathit  Z(r)=\frac{1}{\sqrt{-g_{tt}}}-1=\frac{1}{\sqrt{e^{s_0 (r/L_s)^2+s_1}}}-1}.
\end{equation}

The pulsar's redshift is represented by altering the model's parameter $\xi_1$ in Fig.~\ref{Fig:Matching}\subref{fig:redshift}. When $\xi_1=0$ (general relativity case), then $Z$ corresponding to the core is $Z(0)\approx 0.806$, while i  at the boundary yields $Z_s=Z_{L_s}\approx 0.4$. When considering $\xi_1=-0.03$, then then $Z(0)\approx 0.772$, which is less than that in the general relativity case. As $Z$ approaches the boundary of the pulsar, then  $Z_s\approx 0.397$, that is consistent with the case in general relativity. It is essential to stress that such value is considerably below the higher limit of $Z_s=2$, as shown in Refs.~\cite{Buchdahl:1959zz,Ivanov:2002xf,Barraco:2003jq,Boehmer:2006ye}. Similarly, as $\xi_1=0.03$, the maximum redshift that we can observe at the central will be around $Z(0)\approx 0.834$, exceeding the value predicted by general relativity. The highest redshift decreases as we get closer to the boundary, with $Z_s\approx 0.401$, which is in line with the prediction of general relativity. Interestingly, in the two cases discussed above, the redshift patterns in the $f({\mathcal{Q}})$ theory, specifically in the quadratic form, following the stability guidelines. This indicates that $Z$ remains constant, positive value in the entire region of the pulsar, gradually decreasing towards the boundary, as shown in Fig.~\ref{Fig:Matching}\subref{fig:redshift}.

\subsection{Behaviors of density and pressure}\label{Sec:matt}

\begin{figure*}
\centering
\subfigure[~Density]{\label{fig:density}\includegraphics[scale=0.3]{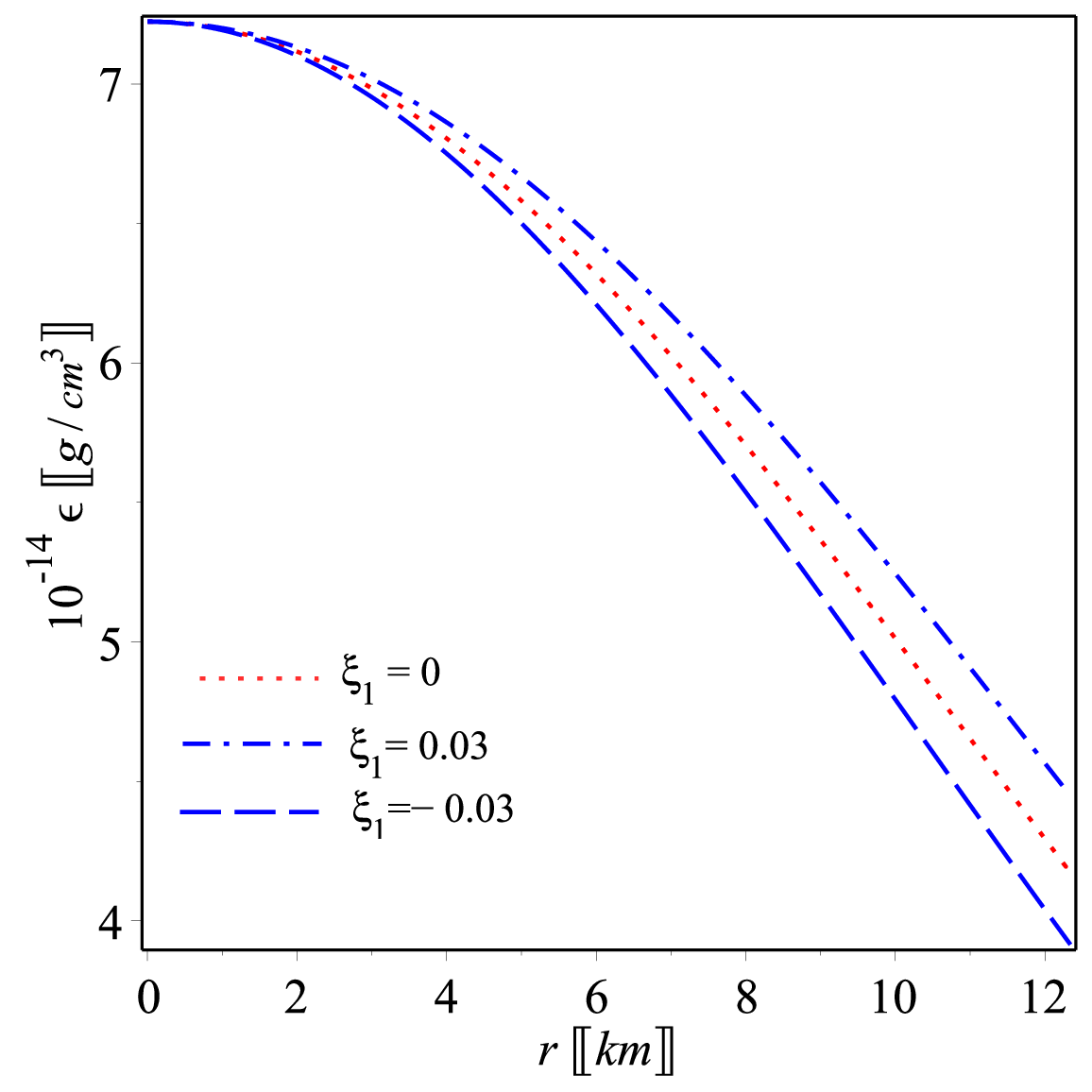}}\hspace{0.5cm}
\subfigure[~Pressure in  radial direction]{\label{fig:radpressure}\includegraphics[scale=0.3]{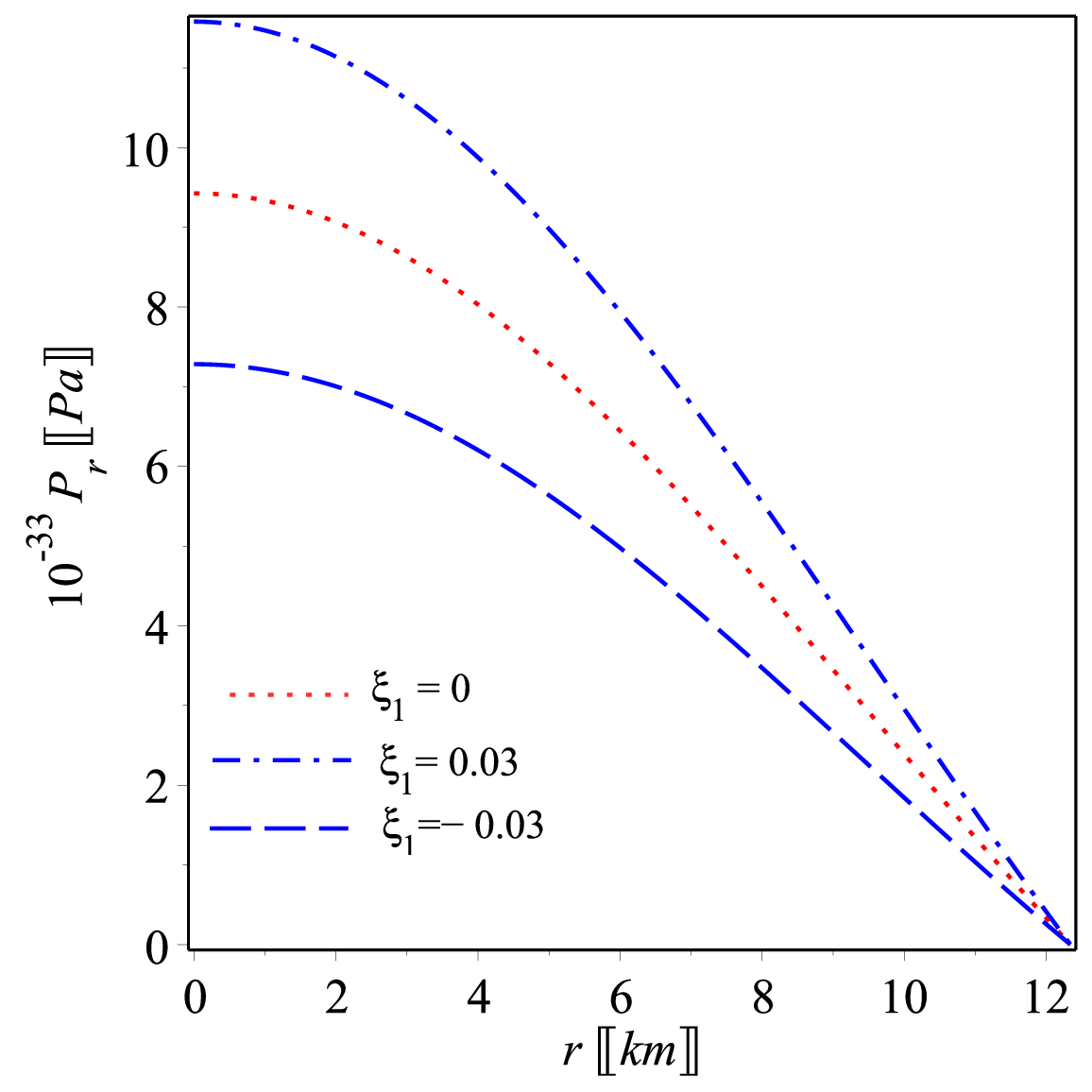}}\\
\subfigure[~Pressure in   tangential direction]{\label{fig:tangpressure}\includegraphics[scale=0.3]{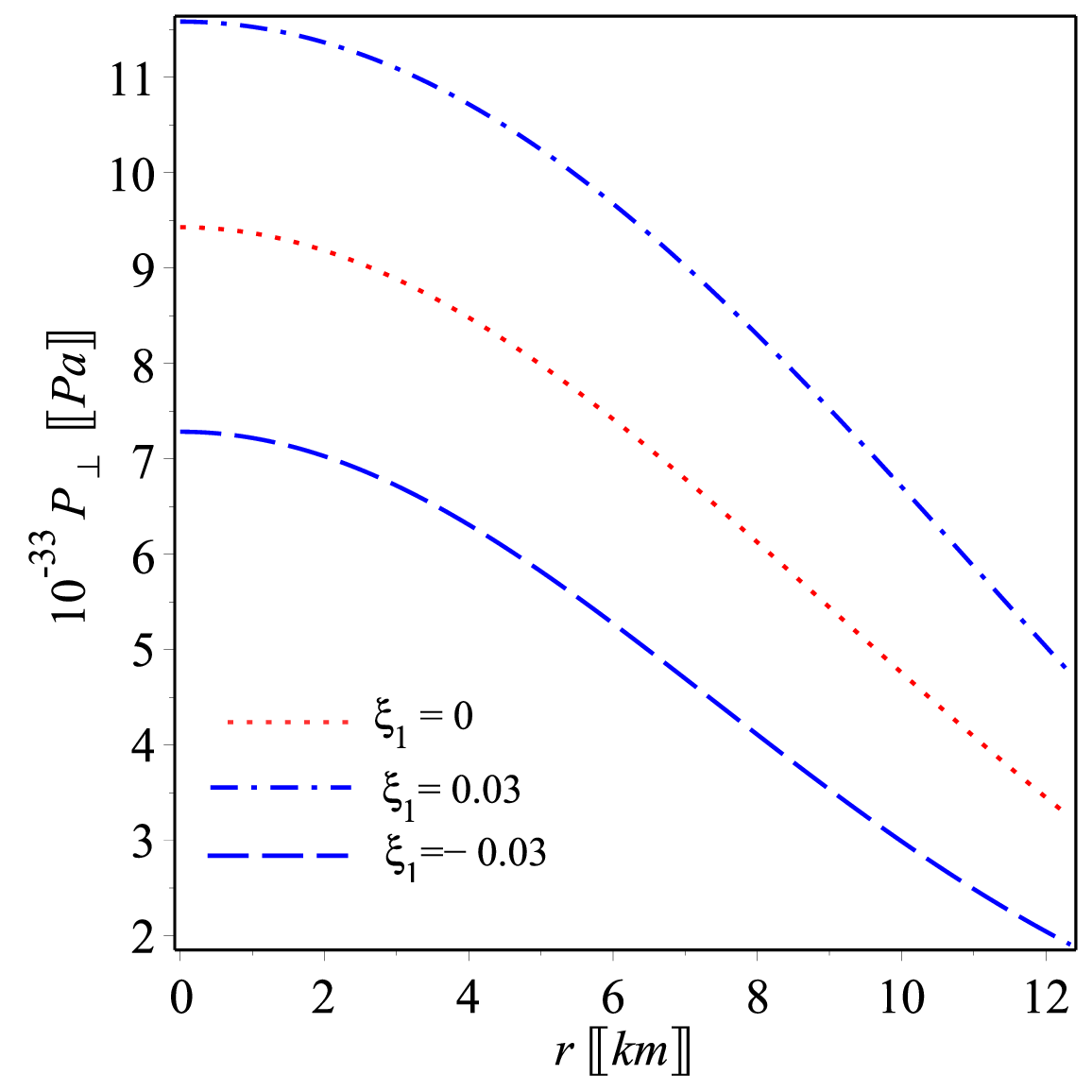}}\hspace{0.5cm}
\subfigure[~Anisotropy]{\label{fig:anisotf}\includegraphics[scale=0.3]{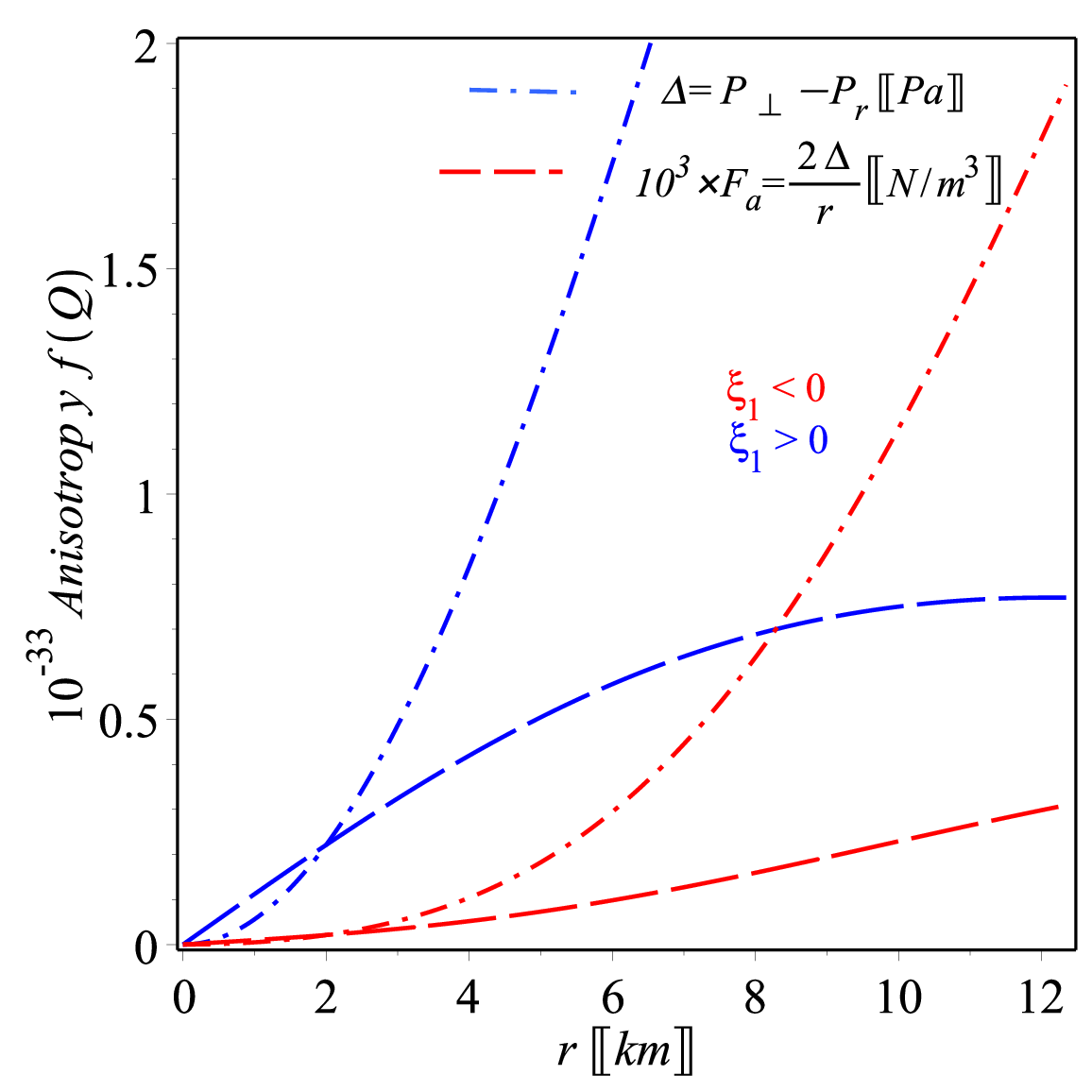}}
\caption{The properties of $\epsilon$, $ P_r$ and $ P_\perp$ for the star ${\mathcal J0740+6620}$, are shown in \subref{fig:density} to \subref{fig:tangpressure}. These are the case for: $\xi_1=0,~\pm0.03$. It is shown how the density and pressures stay in limits of the inner part of a pulsar, reducing as we get closer to the surface. Fig.~\ref{fig:anisotf} illustrates the distribution of the anisotropy , $\Delta(r)=P_\perp-P_r$, in the pulsar for three scenarios: $\xi_1=0,~\pm 0.03$. The lack of anisotropy is evident at the center where $P_\perp=P_r$. However, far from the core, $\Delta>0$ (signifying notable anisotropy with $P_\perp>P_r$) leads to a repulsive force which plays a vital key in controlling the size of the start.}
\label{Fig:dens_press}
\end{figure*}

By examining equation presented in Appendix~\ref{Sec:Appendix New B.} and the values of the parameters obtained in Subsection  \ref{Sec:obs_const}, we can create the plots  of $\epsilon$,  $ P_r$ and $ P_\perp$ as functions $r$, as demonstrated in Fig.~\ref{Fig:dens_press}\subref{fig:density}--\subref{fig:tangpressure}. It is apparent that the profiles for pressure and density comply with the stability criteria relevant to this research area. Their values peak at the center, stay positive without singularities inside the star, and steadily drop towards the boundary, satisfying the conditions $\epsilon(r=0)>0$, $\epsilon'(r=0)=0$, $\epsilon''(r=0)<0$, $\epsilon(r)>0$, $\epsilon'(r)>0$ for the components $ P_r$ and $ P_\perp$. Moreover, we visually depict   $\Delta(r)$, which is figured as $ P_\perp-P_r$, in Fig.~\ref{Fig:dens_press}\subref{fig:anisotf}. It is seen that the anisotropy meets the criteria of the stability by decreasing to zero in the center and increasing towards the face of a star. In situations like the one in our study, where there is pronounced anisotropy, it is important to highlight that an extra force with positive value  plays a significant role in achieving hydrodynamic balance. This force counteracts gravity and is essential for the star's expansion, allowing it to gather more mass than in cases with homogeneous or less extreme conditions. Further elaboration on this topic will be given in Subsection \ref{Sec:TOV}.

Supplying specific parameter values for the pulsar ${\mathcal J0740+6620}$ is crucial according to the model presented in this study. Such numerical values can be summarized as follows. For $\xi_1=\pm 0.03$, the core density is about ${\epsilon_\text{core}\approx 7.22\times 10^{14}}$ g/cm$^{3}$. This corresponds to  $\approx 2.7\times\epsilon_\text{nuc}$, and $P_{r\text{(core)}}$, is about $11.6 \times 10^{33}$ dyn/cm$^2$, which is roughly the same as the center of $P_{\perp\text{(core)}}$. At the outer layer of the pulsar's $\epsilon$ is about $4.45 \times 10^{14}g/cm^3$, which is $1.7\time\epsilon_\text{nuc}$. The surface radial pressure, $P_{r(L_s)}$, is almost zero dyn/cm$^2$, while the surface tangential pressure, $P_{\perp(L_s)}$, is approximately $4.75 \times 10^{33}$ dyn/cm$^2$.

When $\xi_1=-0.03$, the central density is roughly equal to ${\epsilon_\text{core}\approx 7.22\times 10^{14}}$ g/cm$^{3}$, which $\approx 2.7 t\epsilon_\text{nuc}$, and $P_{r\text{(core)}}$, is $\approx 11.6\times 10^{33} \mathrm{dyn}/\mathrm{cm}^2 \approx P_{\perp\text{(core)}}$. The density of the pulsar's at the boundary is around ${\epsilon_s\approx 3.88\times 10^{14}} \mathrm{g}/\mathrm{cm}^{3}\sim 1.4$ times higher than $\epsilon_\text{nuc}$. At the boundary $P_{r\text{(core)}}$ is close to zero dyn/cm$^2$, whereas $P_{\perp\text{(core)}}$ is about $1.85\times 10^{33}$ dyn/cm$^2$. Under these circumstances, the pulsar considers the probability  that at the center of ${\mathcal J0740+6620}$ might be composed of neutrons.  Additionally, the recorded pressure and density readings strongly back up the existence of $\Delta$ of the pulsar.

As outlined in Sec.~\ref{Sec:Model}, a utilization of the KB ansatz given by Eq.~\eqref{eq:KB} to solve the equations \eqref{Rho1}--\eqref{pt1} rather than using the EoS. However, we show that the KB assumption creates a connection between density and pressure. As a result, we introduce a parameter $\varepsilon:=r/L_s$ and proceed to simplify the set of equations presented in Appendix~\ref{Sec:Appendix New B.} by ignoring terms of $O(\varepsilon^2)$ and above. These equations lead to the following relationships
\begin{equation}\label{eq:KB_EoS}
    P_r(\epsilon)\approx b_1 \epsilon+b_2\,, \qquad  p_t(\epsilon) \approx b_3 \epsilon+b_4\,.
\end{equation}
The outcomes are entirely determined by the values of the constants $b_1$, $b_2$, $b_3$, and $b_4$, which are the parameters of the model described in Appendix~\ref{Sec:App_1.}. We now rewrite equations in \eqref{eq:KB_EoS} in a manner that is more accessible from a physical standpoint as
\begin{equation}\label{eq:KB_EoS2}
    P_r(\epsilon)\approx v_r^2(\epsilon-\epsilon_{i})\,, \qquad   P_\perp(\epsilon) \approx v_\perp^2 (\epsilon-\epsilon_{ii})\,.
\end{equation}

In this case, the quantities are presented in a clearer manner: the speed of sound in the radial direction is equal to $v_r^2=b_1$, the density is $\epsilon_i=-b_2/b_1$, the tangential sound speed is $v_\perp^2=b_3$, and the density is $\epsilon_{ii}=-b_4/c_3$. It is important to mention that the density $\epsilon_i$ is equal to the boundary density $\epsilon_s$ and satisfies the boundary condition $P_r(\epsilon_s)=0$. This does not apply to $\epsilon_{ii}$ as $P_\perp$ on the surface may not be zero. These connections include specific examples like the  EoS  for the most compact hadronic matter, where $v_r^2=c^2$ and the EoS of quark matters based on the MIT bag model with $v_r^2=c^2/3$. The speed of sound and surface density are not subject to random selection but are fully determined by model under consideration, as shown in Appendix~\ref{Sec:Appendix New A.}. For $\xi_1=0.03$, using  equations presented  ({\color{blue} \bf C}), we find $v_r^2=b_1\approx 0.50c^2$, $v_\perp^2=b_3\approx 0.26 c^2$, $\epsilon_i=\epsilon_s=-b_2/b_1\approx 4.8\times 10^{14}$ g/cm$^3$ and $\epsilon_{ii}=-b_4/b_3\approx 2.3\times 10^{14}$ g/cm$^3$. Similarly, when $\xi_1=-0.03$, we find that $v_r^2$ is approximately $0.29c^2$, $v_\perp^2$ is approximately $0.2c^2$, $\epsilon_i=\epsilon_s$ is around $3.5\times 10^{14}$ g/cm$^3$, and $\epsilon_{ii}$ is roughly $2.9 \times 10^{14}$ g/cm$^3$. Although the derived equations of state \eqref{eq:KB_EoS2} are mainly applicable to the core part of the pulsar  it is important to mention that te values of $\epsilon$ at the boundary for the two cases with varying $\xi_1$ match closely with the exact values obtained. This affirms the trustworthiness of these equations of state within the star's entire interior, where $0\leq \varepsilon <1$. The examination of the validity of the EoS as well as the speed of sound will be addressed in Subsection \ref{Sec:causality}.

\subsection{Stability conditions of compact objects}
Zeldovich explained in Ref.~\cite{1971reas.book.....Z} that for a star to remain stable, the pressure in the core must not surpass the core density, as discussed in Subsection~\ref{Sec:causality}. This means that
\begin{equation}\label{eq:Zel}
    {\frac{{P}_r(0)}{c^2{\epsilon}(0)}\leq 1.}
\end{equation}
By using the equation presented in Appendix~\ref{Sec:Appendix New A.}, the density and pressure are found in the radial direction value as $r\to 0$
\begin{align}
  {\epsilon}_{_{r\to 0}}={\frac {3s_2 }{{\kappa}{c}^{2}{L_s}^{2}}}\,, \qquad \qquad {P_r}_{_{r\to0}} ={\frac {2s_0-s_2}{{
\kappa}{L_s}^{2}}}.\end{align}
Using the values of  ${\mathcal J0740+6620}$  that were previously acquired in Subsection \ref{Sec:obs_const}, when $\xi_1=0.03$, the Zeldovich inequality \eqref{eq:Zel} translates to $\frac{{P}_r(0)}{c^2{\epsilon}(0)}=0.178$, which is lower than 1. Likewise, when $\xi_1=-0.03$, the inequality gives $\frac{{P}_r(0)}{c^2{\epsilon}(0)}=0.112$, which is below 1 as well. This confirms that the Zeldovich criterion is met in both situations.

\subsection{Mass and radius}\label{Sec:obs_const}

We use the accurate data to determine the radius and mass of ${\mathcal PSR J0740+6620}$, with both parameters taking the specific forms as $M=2.07 \pm 0.11 M_\odot$   $L_s=12.34^{+1.89}_{-1.67}$ km  \cite{Legred:2021hdx}. Such values use data from both NICER and XMM to constraint $\xi_1$.
\begin{figure*}
\centering
{\label{Fig:Mass}\includegraphics[scale=0.35]{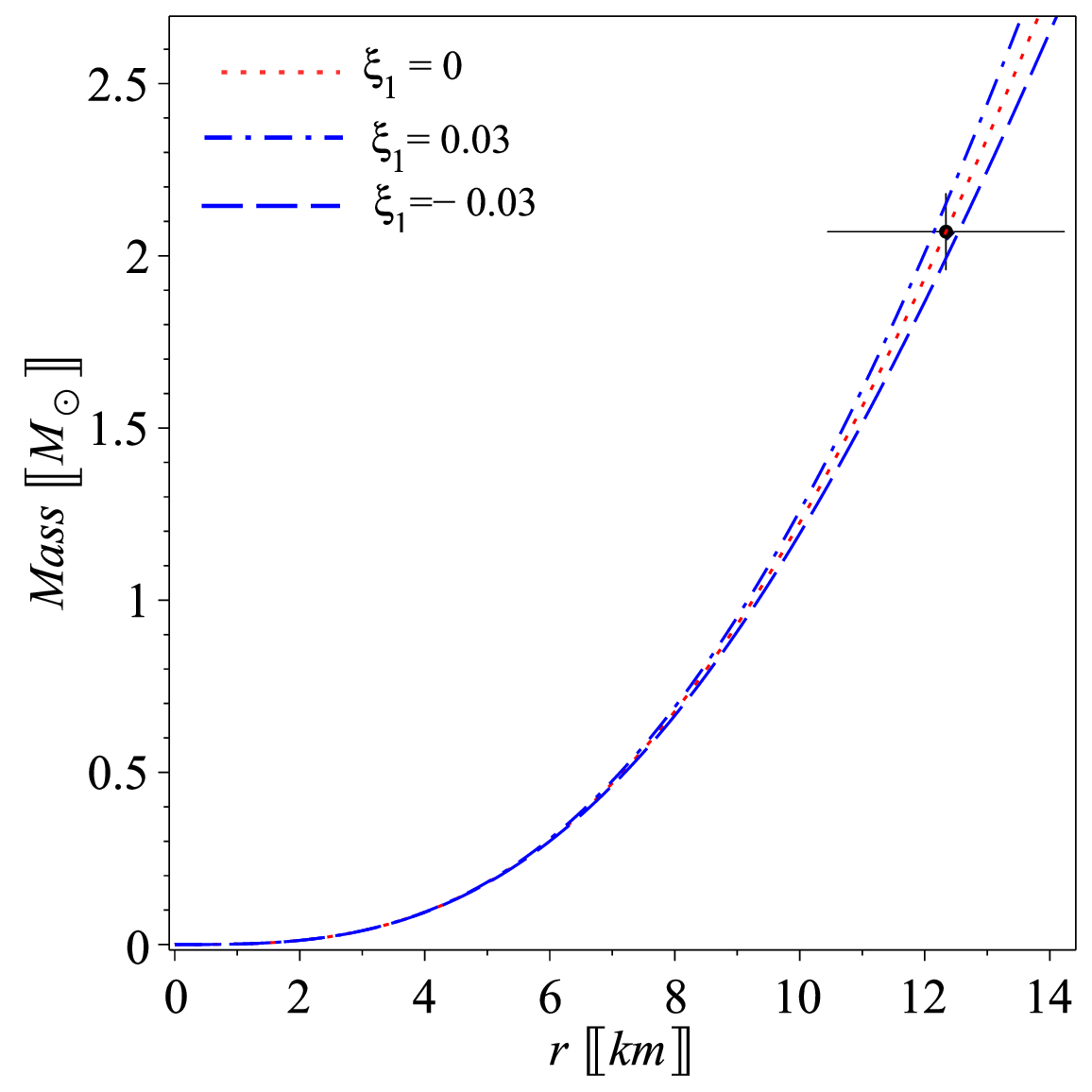}}
\caption{The function of the mass given by Eq.~\eqref{Mf3} for the pulsar ${\mathcal J0740+6620}$ is shown in the limitations imposed by the observations in terms of the radius and mass ($L_s=12.35\pm0.11$ km and $M=2.07\pm 0.11 M_\odot$) in Ref.~\cite{Legred:2021hdx}. If $\xi_1=-0.03$, the parameter set of KB as [ $s_0 \approx0.457$, $s_1\approx-1.142$, $s_2 =0.684$] is used, and when $\xi_1=0.03$, that as [$s_0\approx 0.525$, $s_1 \approx -1.21$, $s_2\approx 0.684$] is taken. For $\xi_1=0$ general relativity, we employ [$s_0 \approx 0.491$, $s_1 =-1.1756$, $s_2 =0.684$]. The plots shows that when $\xi_1$ is greater than zero, the correction derived from the quadratic model tends to underestimate the mass of the pulsar, while it tends to overestimate it when $\xi_1$ is less than zero. It is crucial to emphasize that the same values of parameters are used in all plots throughout this study.}
\label{Fig:Mass1}
\end{figure*}
The function of the mass is described by
\begin{align}\label{Mf3}
{m(r)} =  4\pi\int_{0}^{r} \epsilon(\zeta) \zeta^2 d\zeta \,.
 \end{align}
Using $\epsilon$ given by Eq.~({\color{blue} \bf $\mathrm{B}1$}), for the form  $f({\mathcal{Q}})=Q+\xi {\mathcal{Q}}^2$, we create the plots depicted in Fig.~\ref{Fig:Mass1} for different values of $\xi_1$.
\begin{itemize}
    \item When $\xi_1=0$, which is the case of general relativity, we put  $s_i, i=1..3$ as follows: \{$s_0 \approx 0.491$, $s_1 =-1.1756$, $s_2 =0.684$\}.
    \item With $\xi_1=0.3$, we acquire  mass of the pulsar to be  approximately $2.172 M_\odot$ at a distance of ${L_s}\approx12.11$ km, leading to a compactness of around $C\approx 0.51$. This results in the identification of the constants in the following manner: \{$\xi_1=0.03$, $C=0.51$, $s_0\approx 0.525$, $s_1 \approx -1.21$, $s_2\approx 0.684$\}.
    \item When $\xi_1=-0.03$, the mass of the pulsar is approximately $1.96 M_\odot$ at  ${L_s}\approx12.62$ km, leading to   compactness approximately as
 $C\approx0.486$. These values determine $s_i$ as: \{$\xi_1=-0.03$, $C\approx 0.486$, $s_0 \approx0.457$, $s_1\approx-1.142$, $s_2 =0.684$\}.
\end{itemize}
The model parameter is restricted by these constraints, specifically $0\leq |\xi_1| \leq 300$ km$^2$. Generally, the mass of the pulsar is influenced by the quadratic gravity term of $f{\mathcal{Q}}$ as shown in Fig.~\ref{Fig:Mass1}. If $\xi_1$ is less than 0, the pulsar is expected to have a mass greater than that predicted by general relativity when $\xi_1=0$, given identical value of radius (showing that a small radius for same mass). General relativity states that an object with the same mass will correspond to a larger size.
As a result, this leads to the noted variations of the compactness parameter, as pointed out before. In the following sections, the numerical values shown above in order to evaluate how resilient this model is in the light of the several criteria of the stability.
\subsection{Energy conditions in non-metricity gravity}\label{Sec:Energy-conditions}

It is advantageous to express the equation of motions \eqref{fieldm} as
\begin{equation}\label{eq:fR_MG}
    G_{\mu\nu}=\kappa\left(\mathfrak{T}_{\mu\nu}+\mathfrak{T}_{\mu\nu}^{geom}\right)=\kappa {\mathfrak{T_1}}_{\mu\nu}.
\end{equation}
Here, $G_{\mu\nu}:=\mathcal{R}_{\mu\nu}-g_{\mu\nu}\mathcal{R}/2$ represents the Einstein tensor, which considers the adjustments made by the quadratic form of $f({\mathcal{Q}})$ theory~\cite{DeFelice:2010aj}
\begin{align}
  &  \mathfrak{T}_{\mu\nu}^{geom}=-\frac{1}{\kappa
f_{\mathcal{Q}}}\left(\frac{1}{2} g_{\mu\nu} (f_QQ-f) + 2f_{QQ} P^\lambda{}_{\mu\nu} {\nabla}_\lambda Q \right),
\end{align}
where  ${\mathfrak{T_1}}_{\mu}{^\nu}=diag(-{\epsilon_1} c^2, {P}_r, {P}_\perp, {P}_\perp)$.
\begin{figure*}
\centering
\subfigure[~Radial WEC and NEC ]{\label{fig:Cond1}\includegraphics[scale=0.27]{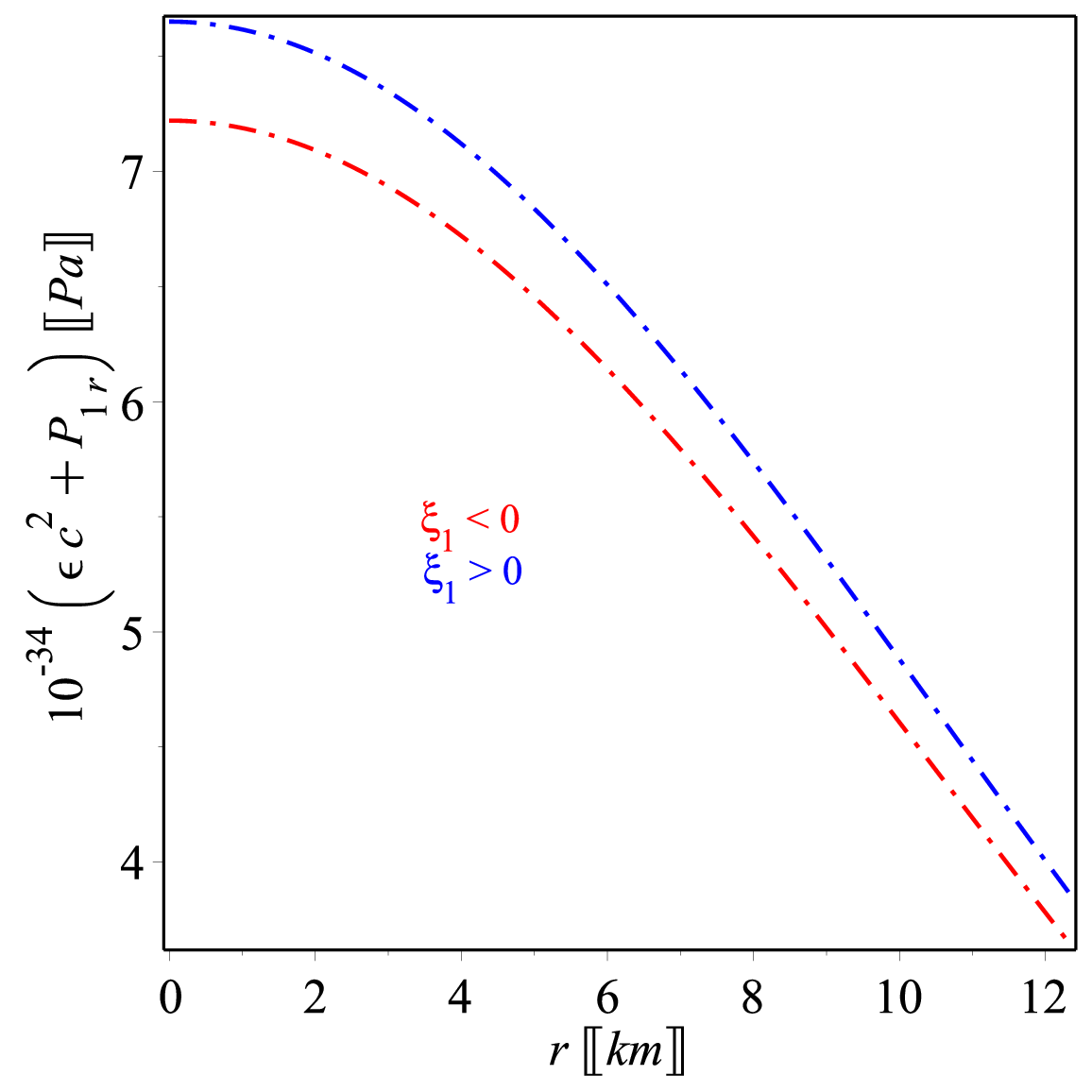}}\hspace{0.2cm}
\subfigure[~Tangential WEC and NEC ]{\label{fig:Cond2}\includegraphics[scale=0.27]{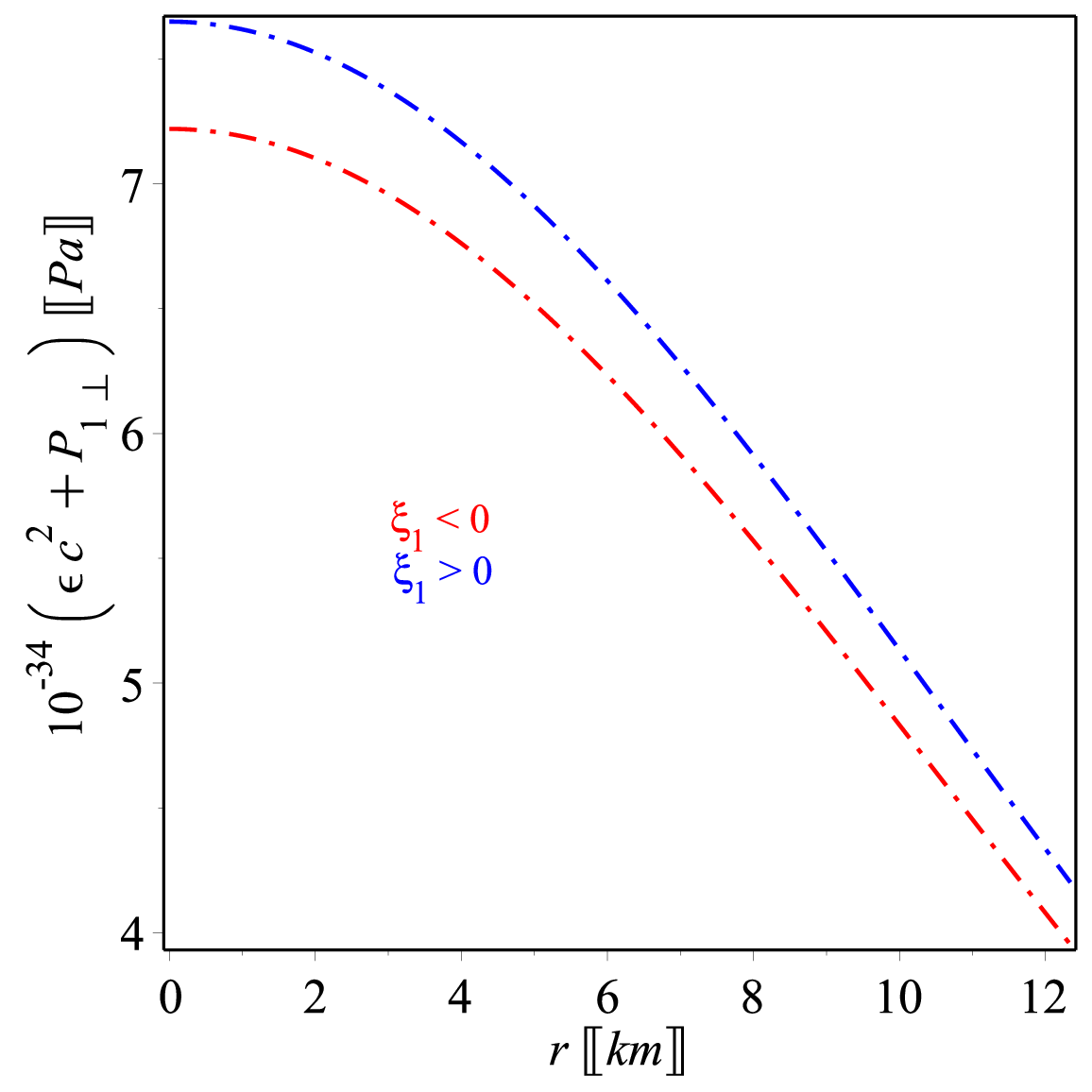}}\hspace{0.2cm}
\subfigure[~The SEC]{\label{fig:Cond3}\includegraphics[scale=.27]{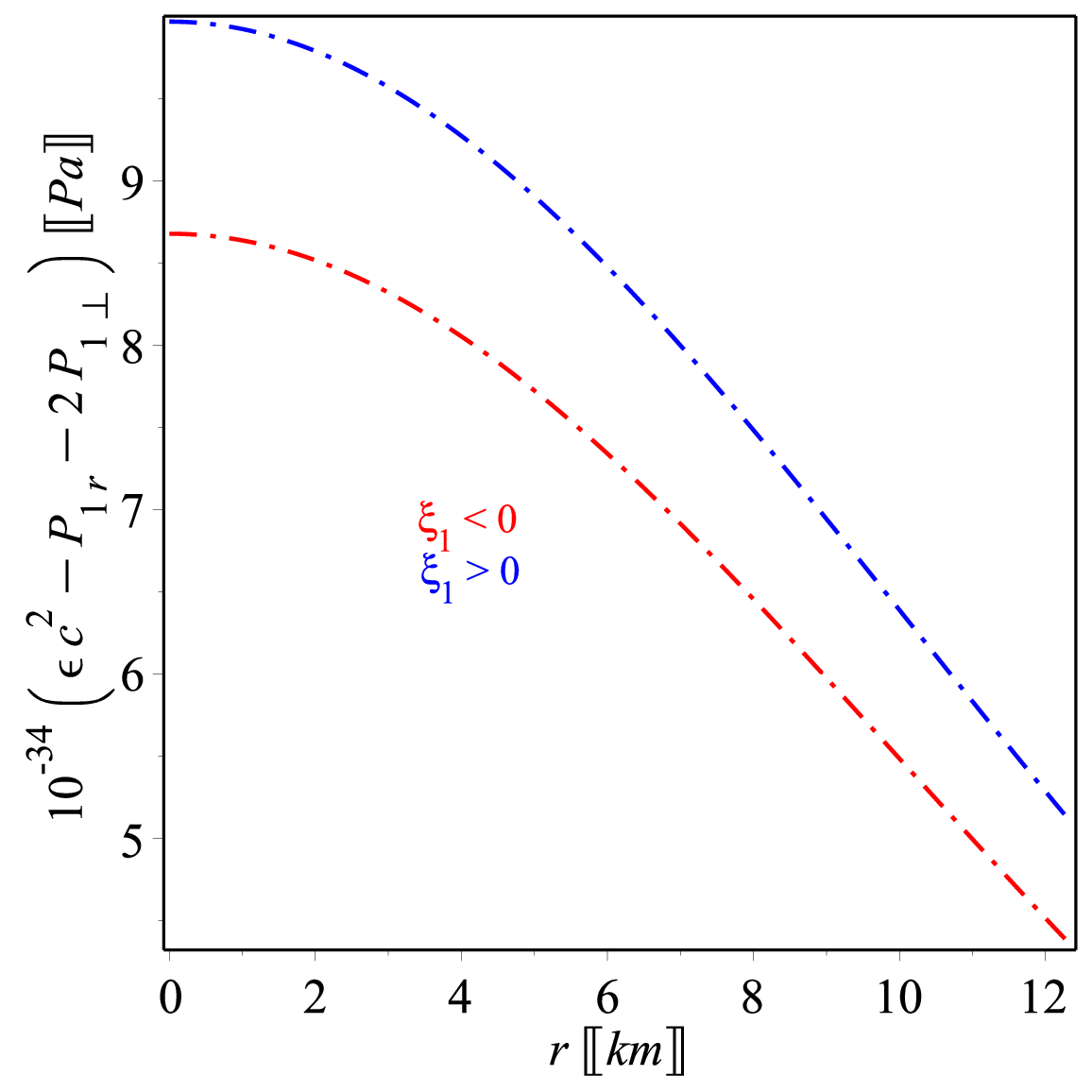}}\\
\subfigure[~ Radial DEC ]{\label{fig:DEC}\includegraphics[scale=.27]{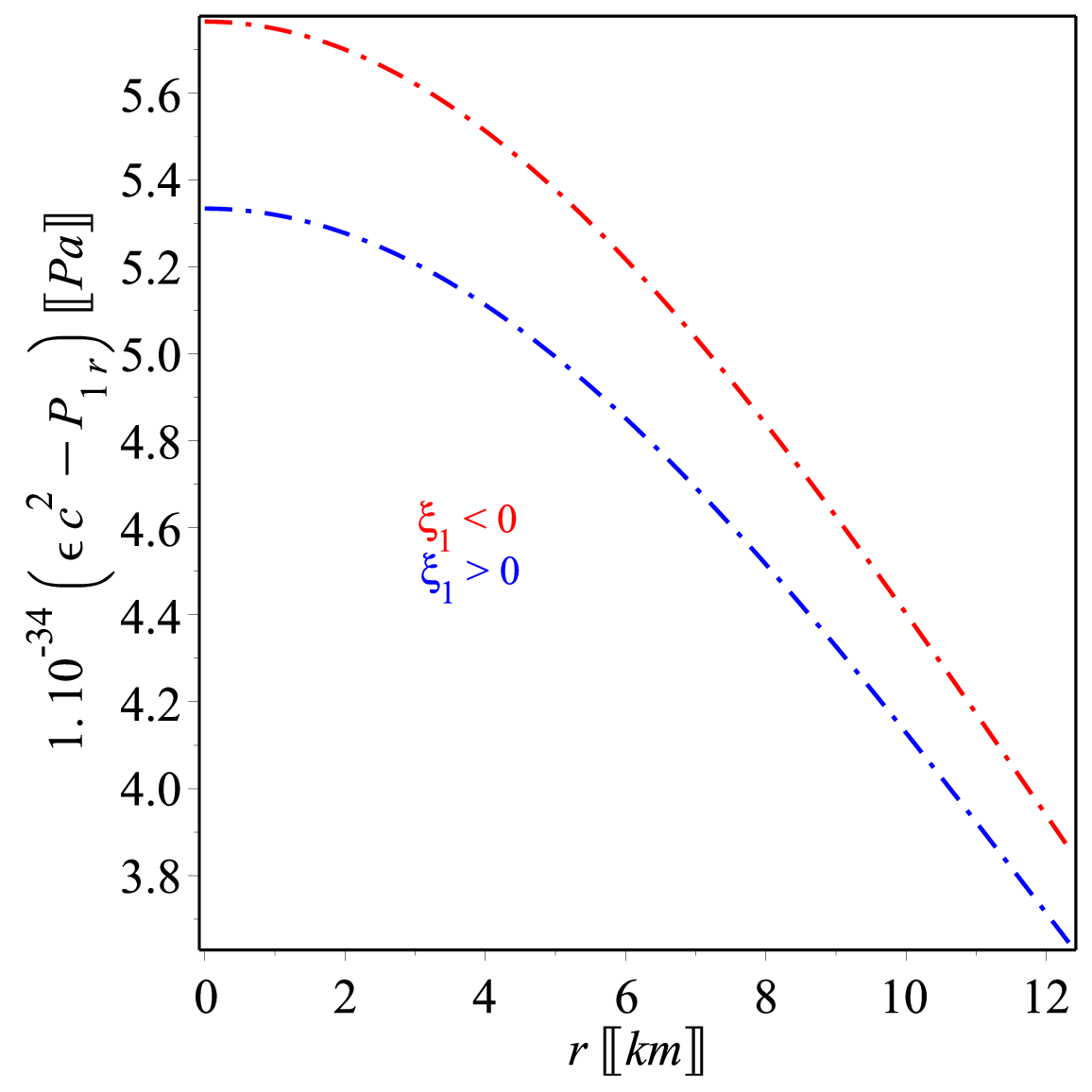}}\hspace{0.2cm}
\subfigure[~Tangential DEC]{\label{fig:DEC1}\includegraphics[scale=.27]{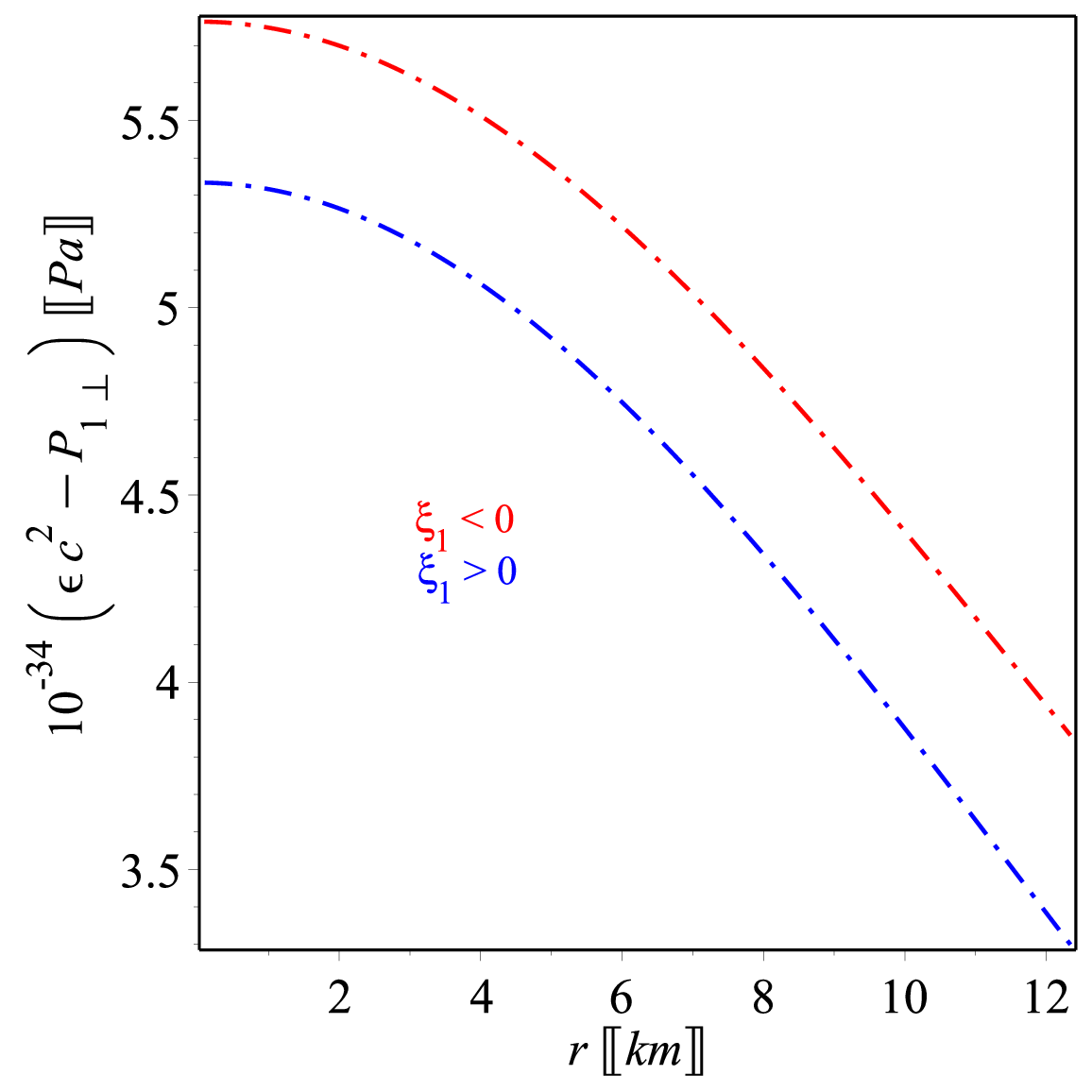}}
\caption{Visual representations that validate the pulsar model ${\mathcal J0740+6620}$ satisfy every energy requirement for the effective matter ${\mathfrak{T_1}}_{\mu\nu}$, as detailed in Subsection \ref{Sec:Energy-conditions}. The situations where $\xi_1>0$ and $\xi_1<0$ correspond to scenarios with $\xi_1=0.03$ and $\xi_1=-0.03$ respectively.}
\label{Fig:EC}
\end{figure*}
By using the focusing theorem and the Raychaudhuri equation, it shows that the tidal tensor's trace should obey these inequalities: $\mathcal{R}_{\mu\nu} u^{\mu} u^{\nu} \geq 0$ and  $\mathcal{R}_{\mu\nu} n^{\mu} n^{\nu} \geq 0$, with $n^{\mu}$, and  $u^{\mu}$, describes the future-directed null vector and a timelike vector respectively (both are arbitrary vectors). In $f(\mathcal{Q})$ gravity, the Ricci tensor is given by $\mathcal{R}_{\mu\nu}=\kappa\left({\mathfrak{T_1}}_{\mu\nu}-\frac{1}{2} g_{\mu\nu} {\mathfrak{T_1}}\right)$, which is an important point to consider. This can be identified by using Eq.~\eqref{eq:fR_MG}. In such case, we can expand the energy requirements to include: $f(\mathcal{Q})$ gravitational theory in the following manner\footnote{Here, the abbreviations WEC $\equiv$ Weak Energy Condition, NEC $\equiv$ Null Energy Condition, SEC $\equiv$ Strong Energy Condition, DEC $\equiv$ Dominant Energy Condition.}.

\begin{itemize}
  \item[a.]  WEC : ${\epsilon_1}\geq 0$, $ {\epsilon_1} c^2+ {P_1}_r > 0$ as well as ${\epsilon_1} c^2+{P_1}_{\perp} > 0$.
  \item[b.]  NEC : ${\epsilon_1} c^2+ {P_1}_r \geq 0$ as well as ${\epsilon_1} c^2+ {P_1}_{\perp} \geq 0$.
  \item[c.]  SEC : ${\epsilon_1} c^2-{P_1}_r-2{P_1}_\perp\geq 0$, ${\epsilon_1} c^2+{P_1}_r \geq 0$ as well as ${\epsilon_1} c^2+{P_1}_{\perp} \geq 0$.
  \item[d.]  DEC :  ${\epsilon_1}\geq 0$, ${\epsilon_1} c^2-{P_1}_r \geq 0$a as well as  ${\epsilon_1} c^2-{P_1}_{\perp} \geq 0$.\\
\end{itemize}
In Fig.~\ref{Fig:EC}, the energy conditions are illustrated in the both cases that the value of $\xi_1$ negative and positive with respect to the total energy-momentum tensor. This data confirms that the present pulsar, i.e., ${\mathcal J0740+6620}$ meets all the requirements of energy.

\subsection{The sound speed in pulsars}\label{Sec:causality}

\begin{figure*}
\centering
\subfigure[~Sound speed for the tangential direction]{\label{fig:vt}\includegraphics[scale=.28]{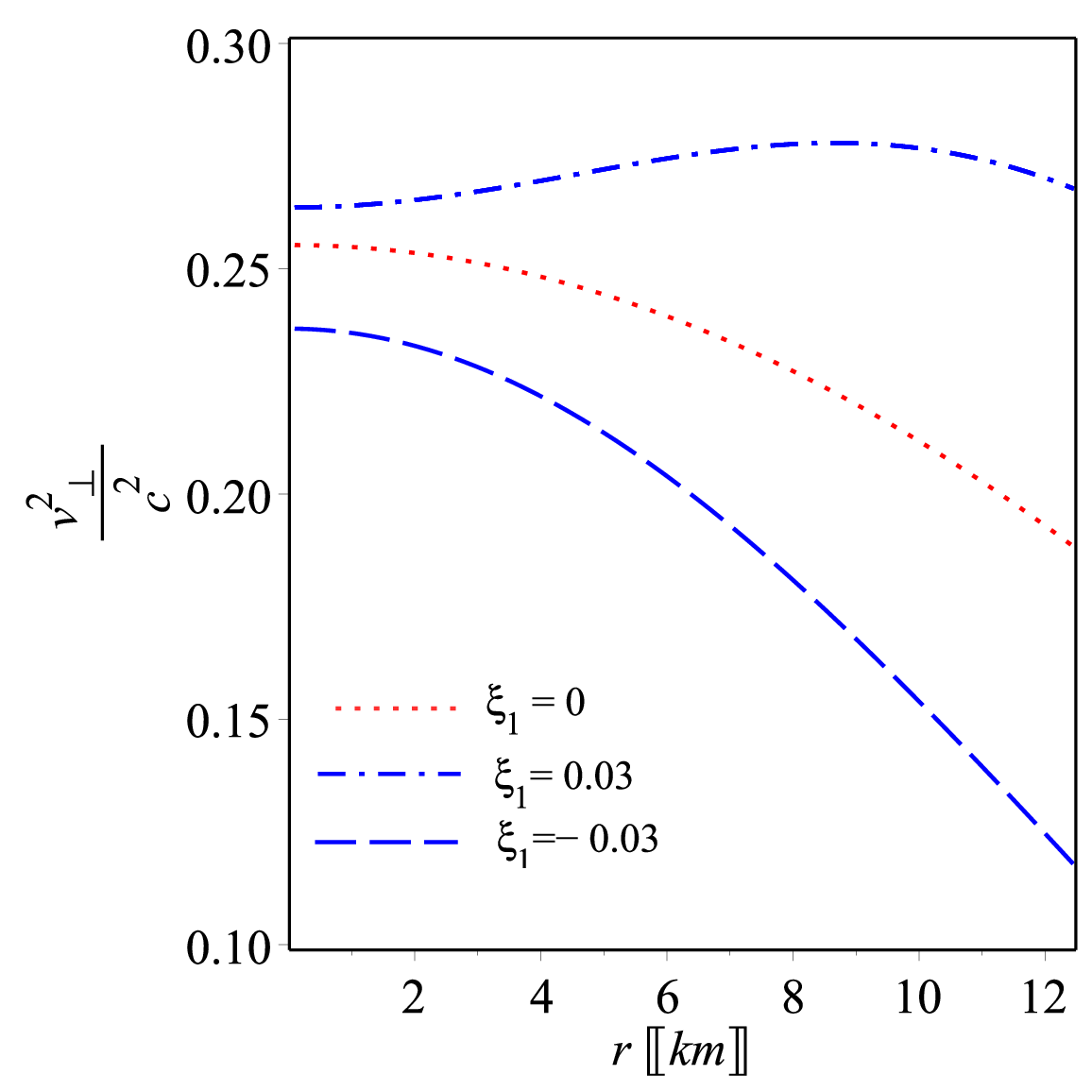}}\hspace{0.2cm}
\subfigure[~Sound speed for the radial direction]{\label{fig:vr}\includegraphics[scale=0.28]{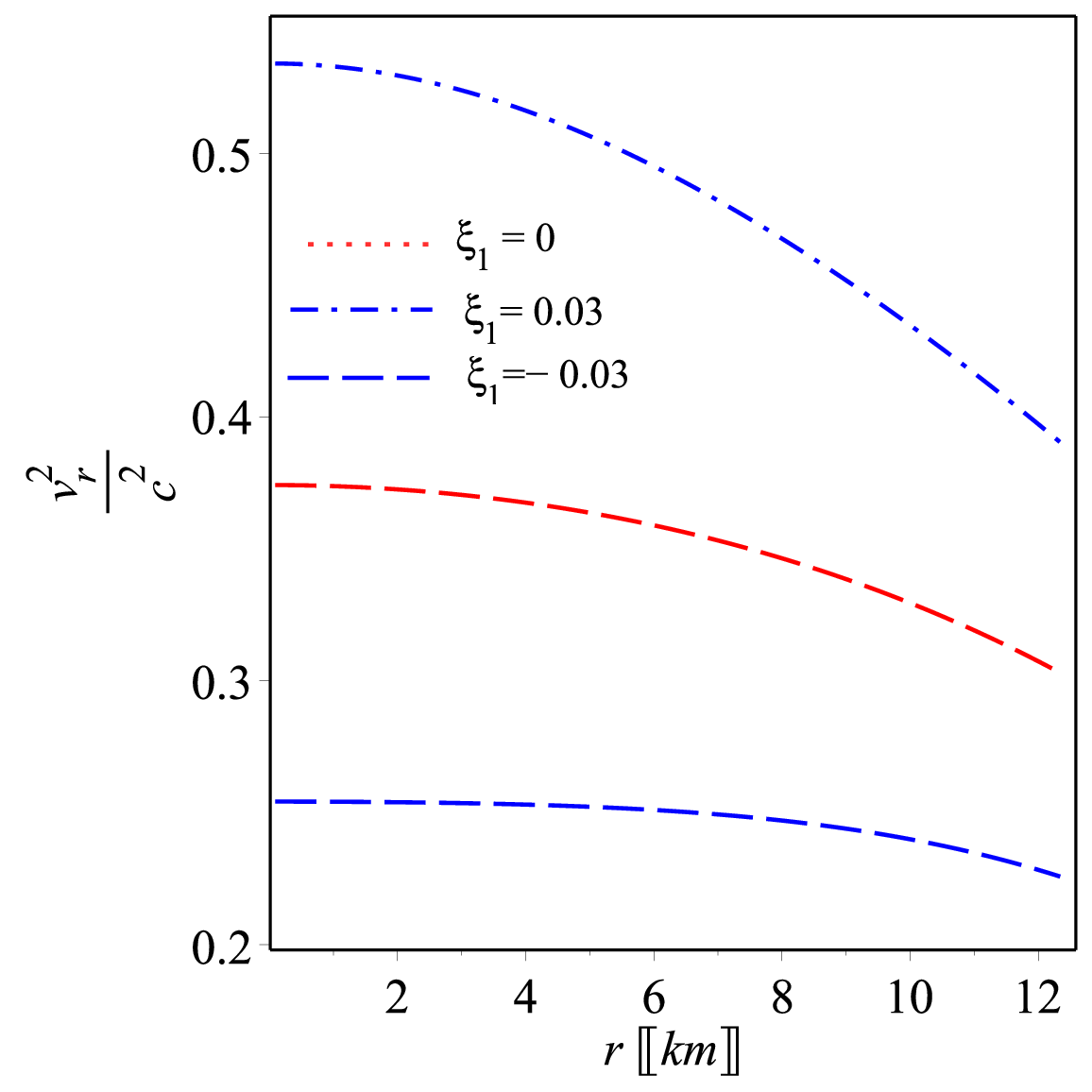}}\hspace{0.2cm}
\subfigure[~Stability in the conditions that the anisotropy is strong]{\label{fig:vt-vr}\includegraphics[scale=.28]{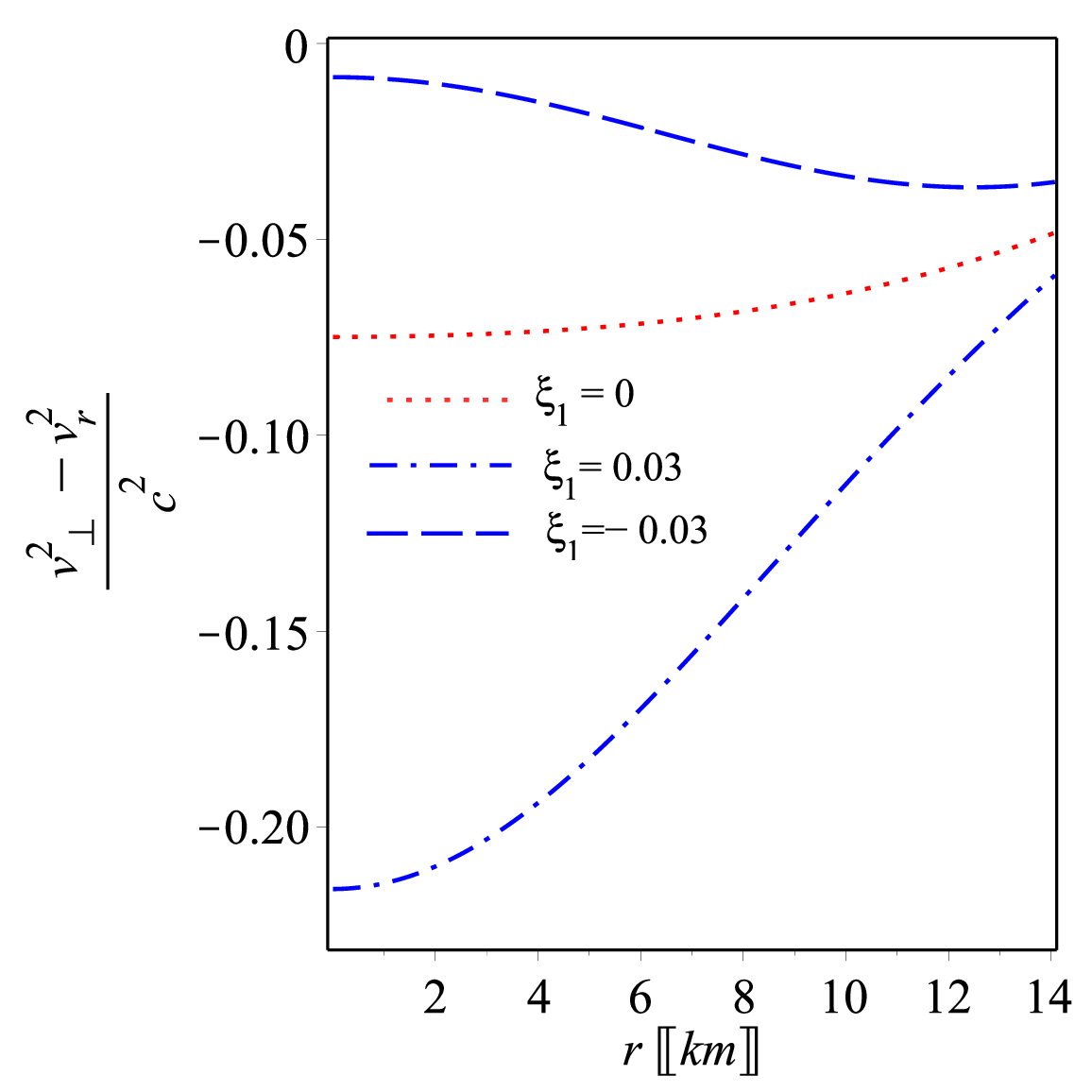}}
\caption{The velocity of sound in the pulsar ${\mathcal J0740+6620}$ is shown for $\xi_1=0,~ \pm 0.03$ in Figs.~\ref{fig:vt} and \ref{fig:vr}, illustrating the propagation of the sound wave in tangential and radial directions based on Eq.~\eqref{eq:sound_speed}. The images in \subref{fig:vt-vr} show that the model meets the stability requirement $(v_\perp^2-v_r^2)/c^2 < 0$ for a highly anisotropic setup.}
\label{Fig:Stability}
\end{figure*}

The requirement of causality is seen as a key characteristic that distinguishes physical systems, stating that the speed of sound cannot exceed the speed of light. When considering the obtained EoS \eqref{eq:KB_EoS2}, the slopes tangential and radial lines show how the speed of sound is characterized as
\begin{equation}\label{eq:sound_speed}
  v_r^2 =  \frac{ d{ P}_r}{d { \epsilon}}=  \frac{P'_r}{{ \epsilon'}}, \quad
  v_ \perp^2 = \frac{d{ P}_ \perp}{d{   \epsilon}}= \frac{P'_ \perp}{{ \epsilon'}}.
\end{equation}
By utilizing equations presented in Appendix~\ref{Sec:Appendix New B.}, we calculate derivatives of the  pressure components and density, that we are mentioned in Appendix~\ref{Sec:App_2.}. We demonstrate how the speed of sound spreads in pulsar ${\mathcal J0740+6620}$ in tangential and radial directions by examining various values of $\xi_1$.
These are shown in Figs.~\ref{Fig:Stability}\subref{fig:vt} and \subref{fig:vr}. It is demonstrated that $v_r^2/c^2$ and $v_\perp^2/c^2$ lie in the ranges of $0\leq {v_r^2}/c^2\leq 1$ and $0\leq {v_\perp^2/c^2} \leq 1$, and thus both the stability and causality conditions can be satisfied. Moreover, it is clear that within pulsar ${\mathcal J0740+6620}$, $-1< (v_\perp^2-v_r^2)/c^2 < 0$ remains true in the interior as shown in Fig.~\ref{Fig:Stability}\subref{fig:vt-vr}.
This condition is necessary for the fact that that a structure of a star, which is not uniform in all directions \cite{Herrera:1992lwz}, is stable.
It should be noted that the sound speed changes in the tangential and in  the radial directions as we move radially, as shown in Figs.~\ref{Fig:Stability}\subref{fig:vt} and \subref{fig:vr}. With $\xi_1=0.03$, it can be seen that the values of $v_\perp^2/c^2$ lie between 0.25 and 0.30, while $v_r^2/c^2$ ranges from 0.40 to 0.55.  When $\xi_1=-0.03$, it is discovered that $v_\perp^2/c^2$ is between 0.1 and 0.25, and $v_r^2/c^2$  is between 0.2 and 0.3. The top values for these intervals align with the sound speed at the center. Notably, these values closely align with the corresponding values found in subsection \ref{Sec:matt} using the EoS \eqref{eq:KB_EoS2}: when ($\xi_1$ = 0.03), we have ($v_r^2 \approx 0.47c^2$) and ($v_\perp^2 \approx 0.28c^2$); and when ($\xi_1$ = -0.03), we get ($v_r^2 \approx 0.25c^2$) and ($v_\perp^2 \approx 0.18c^2$).

It is important to emphasize the following point. In the quadratic theory of $f(\mathcal{Q})$ gravity for a negative value of $\xi_1$, the sound speed for the radial direction, particularly $v_r^2<0.3 c^2$, implies that pulsar ${\mathcal J0740+6620}$'s equation of state is not overly rigid in contrast to the isotropic case of general relativity, where $c_s^2 \sim 0.75 c^2$ in the center \cite{Legred:2021hdx}.
This result is consistent with the predictions from the EoS obtained from the tidal deformability observed in the detection of gravitational waves.
It should be noted, nevertheless, that even under ideal conditions, the radial speed of sound determined by taking the quadratic theory of $f(\mathcal{Q})$ gravity will not become over than the suggested conformal maximum sound speed, which is denoted as $c_s^2=c^2/3$, and is determined through perturbative quantum chromodynamics \cite{Bedaque:2014sqa} in the instance where $\xi_1<0$.

\subsection{ Influences of the quadratic component on the equilibrium conditions and the stability of astrophysical fluid}\label{Sec:TOV}

With the TOV equation, we understand the hydrodynamical equilibrium under the consideration. The corresponding TOV equation to fit a specific $f(\mathcal{Q})$ theory can be expressed as
\begin{equation}\label{eq:RS_TOV}
{\mathrm  F_a}+{\mathrm F_g}+{\mathrm F_h}+{\mathrm F_Q=0}\,.
\end{equation}
Here, ${\mathrm F_g}$, ${\mathrm F_h}$, and  ${\mathrm F_a}$ refer to the usual gravitational, hydrostatic, and anisotropic forces, as well as the additional force ${f(\mathcal{Q})}$ originated from the quadratic term in $f(\mathrm{Q})$ gravity (i.e., the quadratic term in gravity of $\xi_1 \mathcal{Q}^2$). These forces are characterized by
\begin{eqnarray}\label{eq:Forces}
  {\mathrm  F_a} &=&\frac{ 2{\mathrm  \Delta}}{\mathrm r} ,\qquad
 {\mathrm  F_g} = -\frac{{\mathrm  M_g}}{r}({\mathrm  \epsilon c^2}+{\mathrm P_r})e^{\varrho_1/2} ,\nonumber\\
  {\mathrm  F_h} &=&-{\mathrm  P'_\perp} ,\qquad
 {\mathrm F_\mathcal{Q}}= 2\xi_1(\mathcal{Q}')\,.
\end{eqnarray}
where ${\mathrm F_g}$ is the gravitational force. Here, we define $\varrho_1:=a-b$, and the term $\mathrm{M}_g$ figures the gravitational mass of the isolated system in 3-dimensional space ${\mathrm V}$ (when $t$=const).

\begin{figure}
\centering
\subfigure[~The limits of the TOV equation for general relativity]{\label{fig:GR}\includegraphics[scale=0.28]{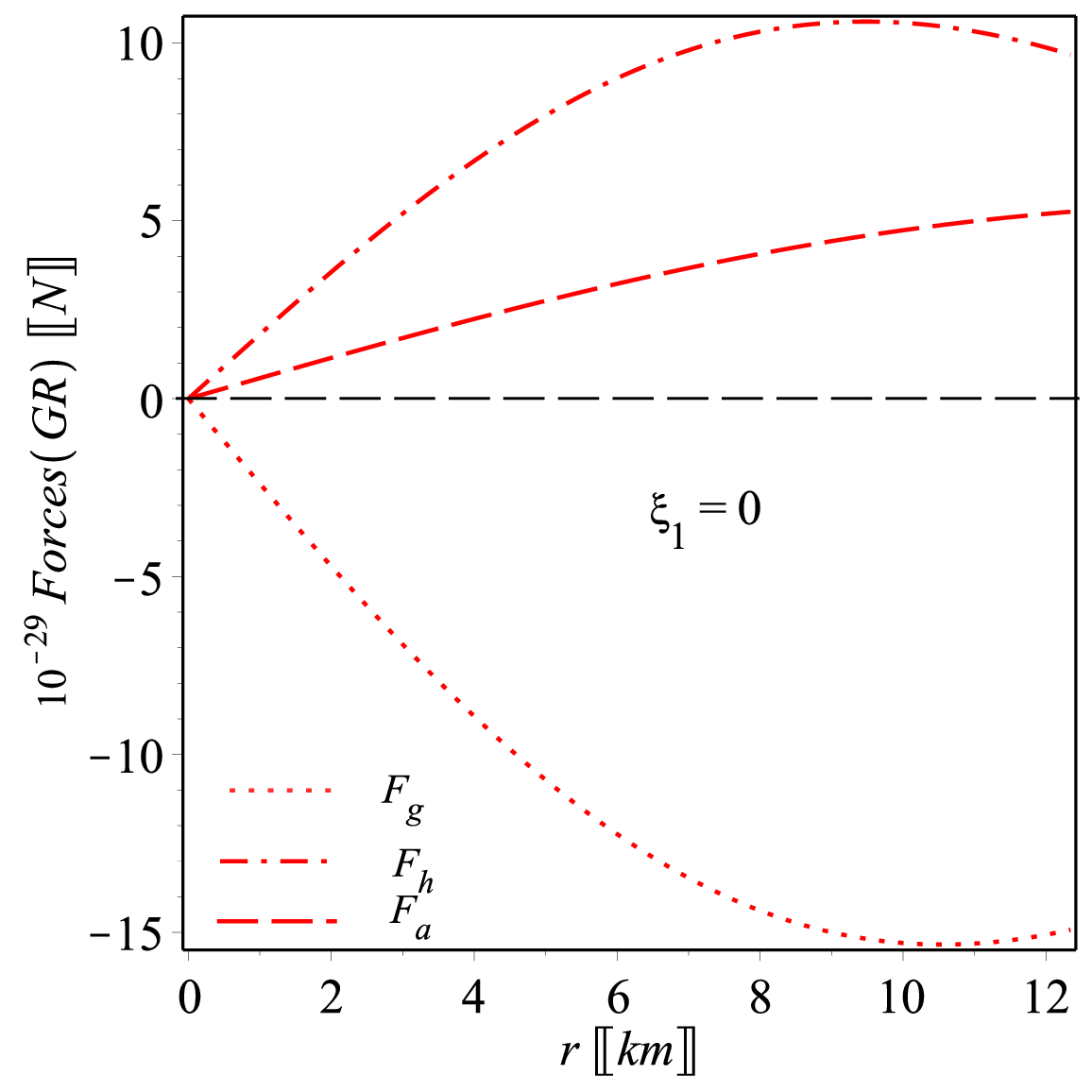}}\hspace{0.2cm}
\subfigure[~The limits of the TOV equation if $\alpha_2>0$ with $\xi_1=0.03$]{\label{fig:ve}\includegraphics[scale=0.28]{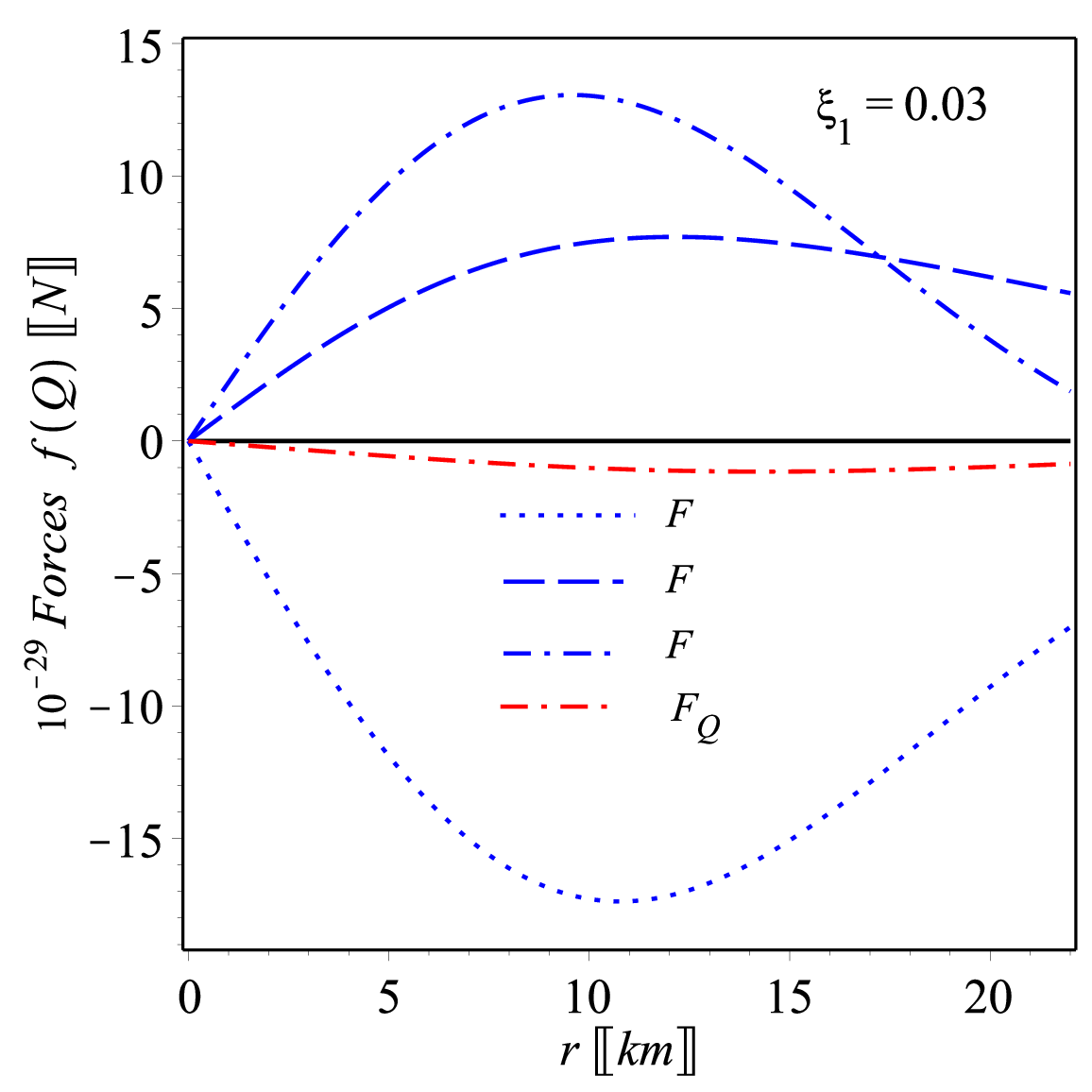}}\hspace{0.2cm}
\subfigure[~The limits of the TOV equation if $\alpha_2>0$ with $\xi_1=-0.03$]{\label{fig:nv}\includegraphics[scale=0.28]{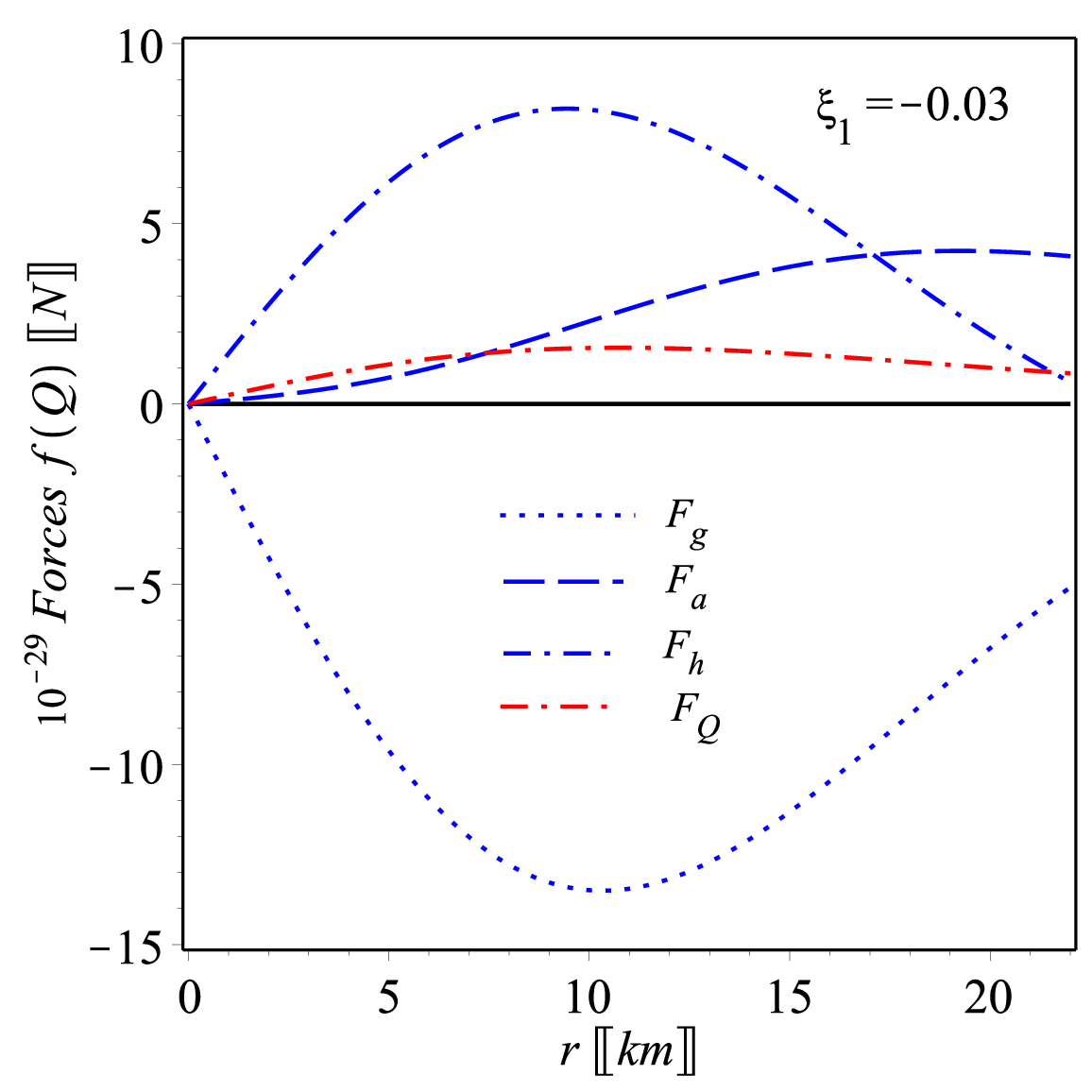}}
\caption{The TOV constraint represented by Eq.~\eqref{eq:RS_TOV}: Various forces acquired from Eq.~\eqref{eq:Forces}, on the pulsar ${\mathcal J0740+6620}$ are shown when $\xi_1=0,~\pm 0.03$. If $\xi_1=0.03$, the additional negative force introduced by the quadratic correction increases the force of gravitational collapse. When $\xi_1=-0.03$, the force for the gravitational collapsing is lessened by the additional positive force introduced from the quadratic correction.}
\label{Fig:TOV}
\end{figure}

The balance of the hydrodynamics is investigated through the TOV equation for a specific case of $f(\mathcal{Q})$ gravity. As a consequence, we find that  ${\mathcal M_g(r)}$ is described by
\begin{eqnarray}\label{eq:grav_mass}
{\mathcal M_g(r)}&=&{\int_{\mathit V}}\Big({{{\mathfrak T_1}}}{^r}{_r}+{{\mathfrak T_1}}{^\theta}{_\theta}+{\mathfrak{T_1}}{^\phi}{_\phi}-{\mathfrak{T_1}}{^t}{_t}\Big)\sqrt{-g}\,dV\nonumber\\
&=&e^{-a}(e^{a/2})'  e^{b/2} r =\frac{1}{2} r a' e^{-\varrho_1/2}\,.
\end{eqnarray}
Consequently, the gravitational force is represented as ${\mathit F_g}=-\frac{s_0 r}{L_s^2}({\mathit \epsilon c^2}+{ P_{1r}})$. By employing the gradients in Eqs.~(\ref{Eq:NewB2}) and (\ref{Eq:NewB3}) and the field equations in~(\ref{Eq:NewB1}), we confirm that the theory of quadratic gravity, expressed as $f(\mathcal{Q})$, adheres with Eq.~\eqref{eq:RS_TOV}. Therefore, our study produces a stable model in the frame of pulsar ${\mathcal J0740+6620}$ without any relation to the values of the parameter $\xi_1$, while also considering the case of general relativity ($\xi_1=0$), depicted in Fig.~\ref{Fig:TOV}.

It is essential to emphasize that the condition of strong anisotropy, represented by $\Delta>0$, results in the development of a force that opposes gravity.  This has a significant effect in supporting more mass as the size of a star increases, while maintaining structural stability.
In Figs. \ref{Fig:TOV}\subref{fig:ve} and \subref{fig:nv}, it is demonstrated that the extra force arising from the quadratic form of $f(\mathcal{Q})$ theory influences the  gravitational collapse. When $\xi_1$ is positive, this force aids in reinforcing the collapse, whereas for negative values of  $\xi_1$ it partially counteracts the collapse. This check is supported  by previous results related to stellar observations ${\mathcal J0740+6620}$ in Subsection \ref{Sec:obs_const}, which are \{$\xi_1=-0.03$, $C\approx 0.486$, ${L_s}\approx12.62$\} and  \{$\xi_1=0.03$, $C=0.51$, ${L_s}\approx12.11$\}.

\begin{figure}
\centering
\subfigure[~The index of adiabatic ]{\label{fig:gamar1}\includegraphics[scale=0.28]{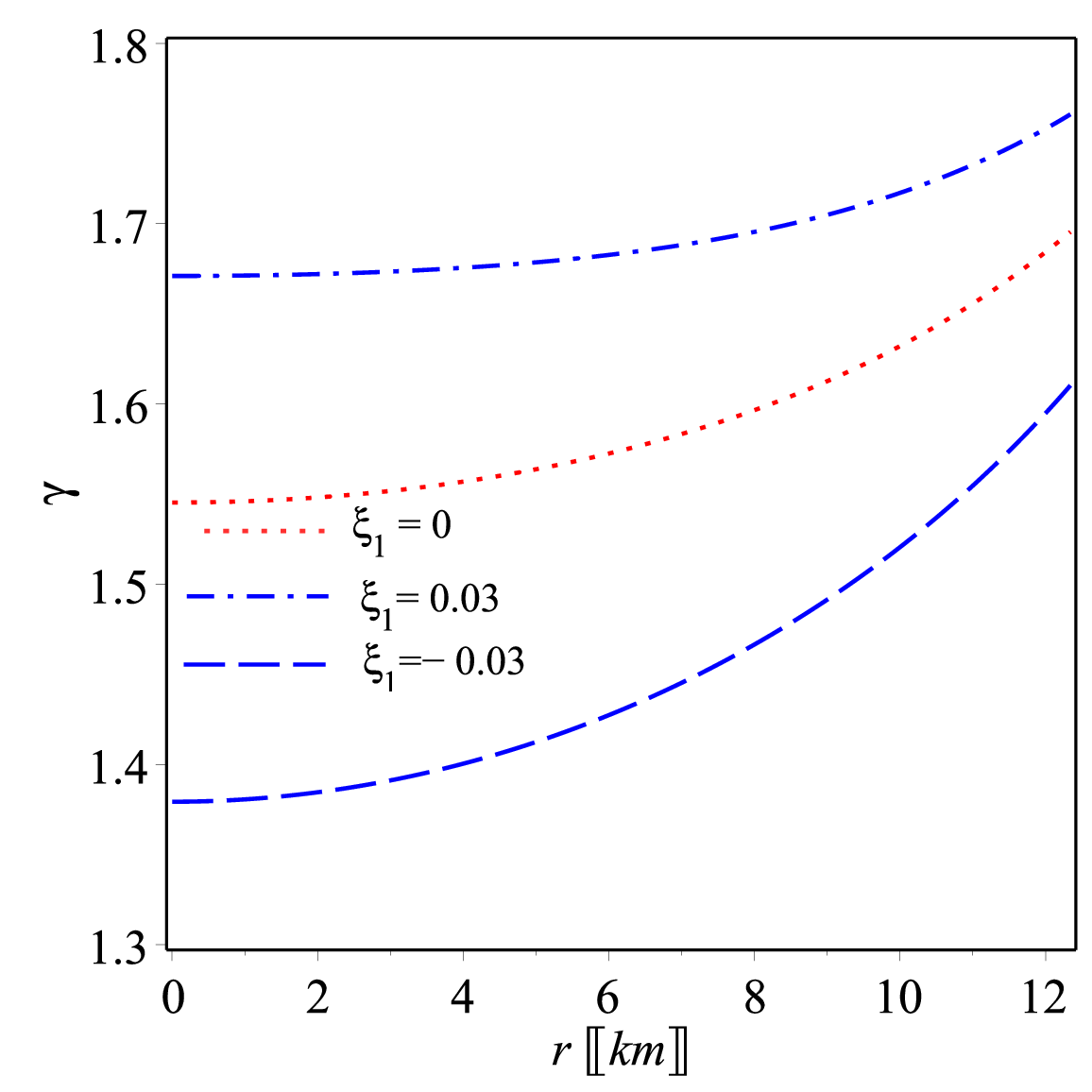}}\hspace{0.2cm}
\subfigure[~The index of adiabatic in the radial direction]{\label{fig:gamar}\includegraphics[scale=0.28]{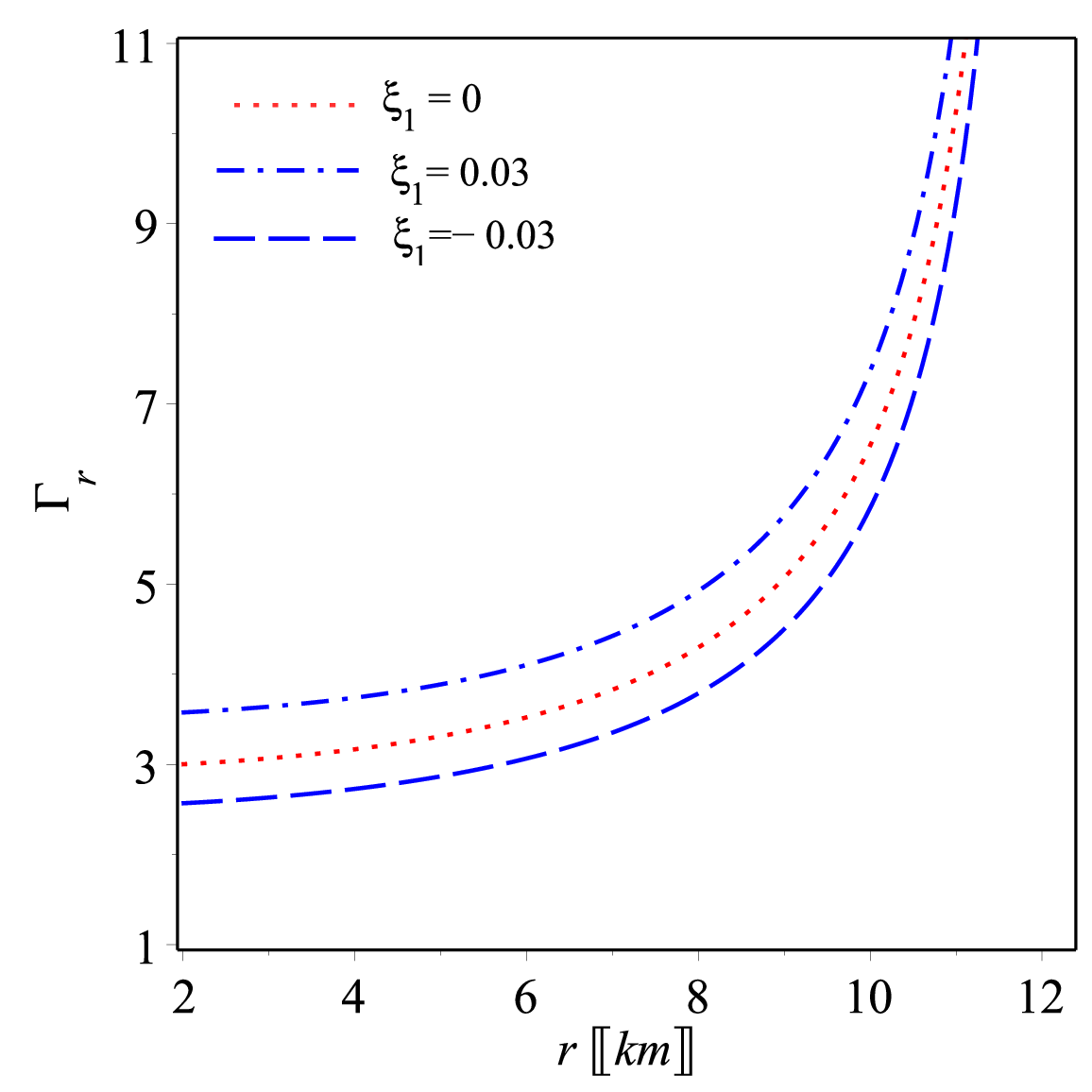}}\hspace{0.2cm}
\subfigure[~The index of adiabatic in the tangential  direction]{\label{fig:gamar2}\includegraphics[scale=0.28]{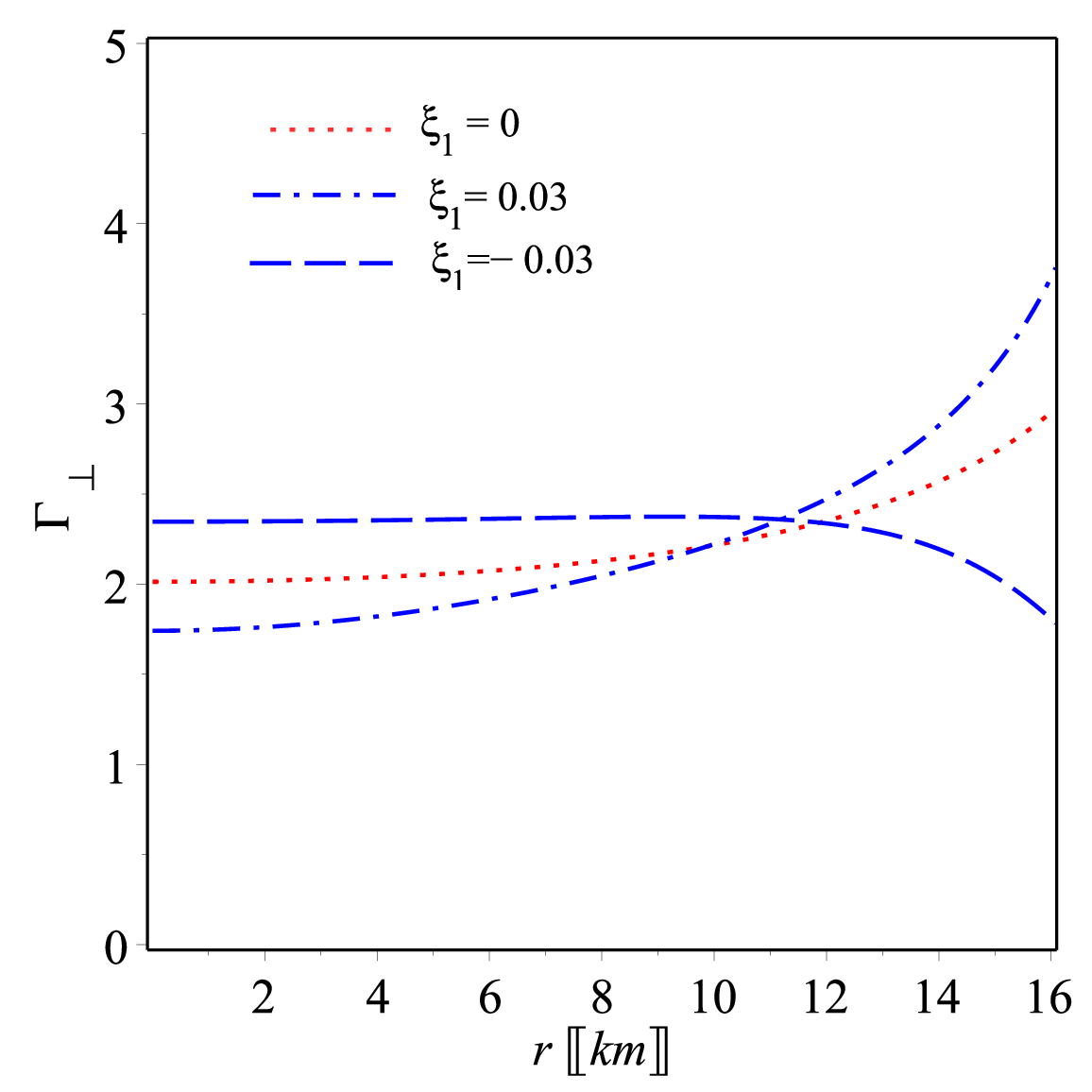}}
\caption{ The plots of \subref{fig:gamar1}, \subref{fig:gamar}, and \subref{fig:gamar2} verify the stability of pulsar ${\mathcal J0740+6620}$, due to $\gamma>4/3$, and show that both $\Gamma_r$ and $\Gamma_t$ are greater than $\gamma$  across the entire of the star. This is a vital condition for a fluid with its strong anisotropy.}
\label{Fig:Adiab}
\end{figure}
%
In Newtonian gravity, if the adiabatic index (a measure of how gases respond to changes in temperature and pressure) exceeds 4/3, there's no maximum mass limit. This is a widely accepted fact in this context. To simplify our discussion, to reach a stable setup under Newton's theory of gravity $\gamma<4/3$.
On the other hand, for a relativistic model of a neutron pulsar with anisotropy, it is shown that the star may retain the stability for the disturbances in the radial direction when $\gamma > 4/3$. Therefore, the adiabatic index is determined as \cite{Chandrasekhar:1964zz,chan1993dynamical}
\begin{equation}\label{eq:adiabatic}
{\gamma}=\frac{4}{3}\left(1+\frac{{ \Delta}}{r|{  P}'_r|}\right)_\mathrm{max},\quad
{\Gamma_r}=\frac{{\epsilon c^2}+{P_r}}{{P_r}}{v_r^2}, \quad
{\Gamma_ \perp}=\frac{{\epsilon c^2}+{P_\perp}}{{p_\perp}}{v_\perp^2 .}
\end{equation}
In the isotropic scenario, i.e., $\Delta=0$, we get $\gamma = 4/3$. When  $\Delta < 0$, we find $\gamma < 4/3$, which is consistent with the case in Newtonian gravity. Contrarstingly, for strong anisotropy, i.e.,  $\Delta > 0$, as in the current investigation, we observe $\gamma > 4/3$. The stability is maintained when $\Gamma$ equals $\gamma$, while a stable equilibrium requires $\Gamma$ to be greater than $\gamma$, as discussed in Refs.~\cite{chan1993dynamical,1975A&A....38...51H}. By applying the field equations (\ref{Eq:NewB1})--(\ref{Eq:NewB3}), we see that the quadratic form of $f(\mathcal{Q})$ gravity can yield a viable anisotropic model in terms of the pulsar ${\mathcal J0740+6620}$, irrespective the values of $\xi_1$, as demonstrated in Fig. \ref{Fig:Adiab}.

\section{Mass-radius relation and properties of compact objects}\label{Sec:EoS_MR}

\begin{figure}[th!]
\subfigure[~EoS for the radial direction]{\label{fig:RfEoSp}\includegraphics[scale=0.45]{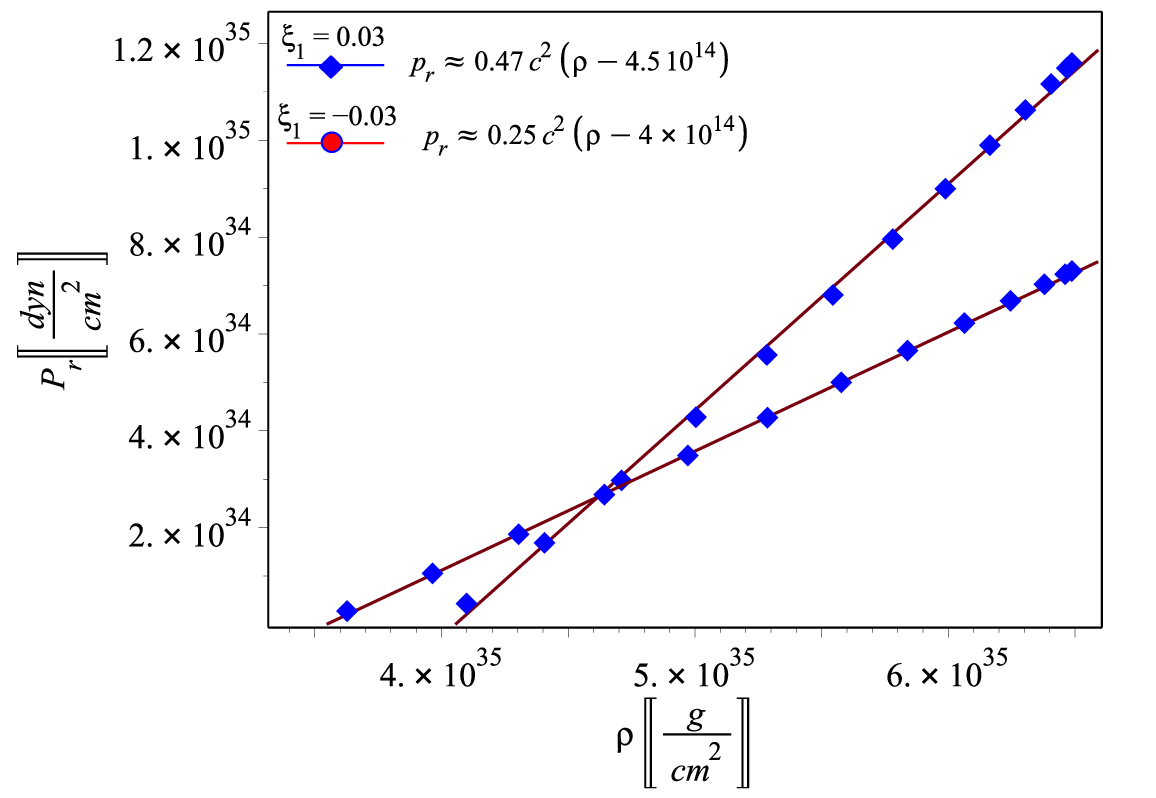}}
\subfigure[~EoS for the tangential direction]{\label{fig:TEoSn}\includegraphics[scale=0.45]{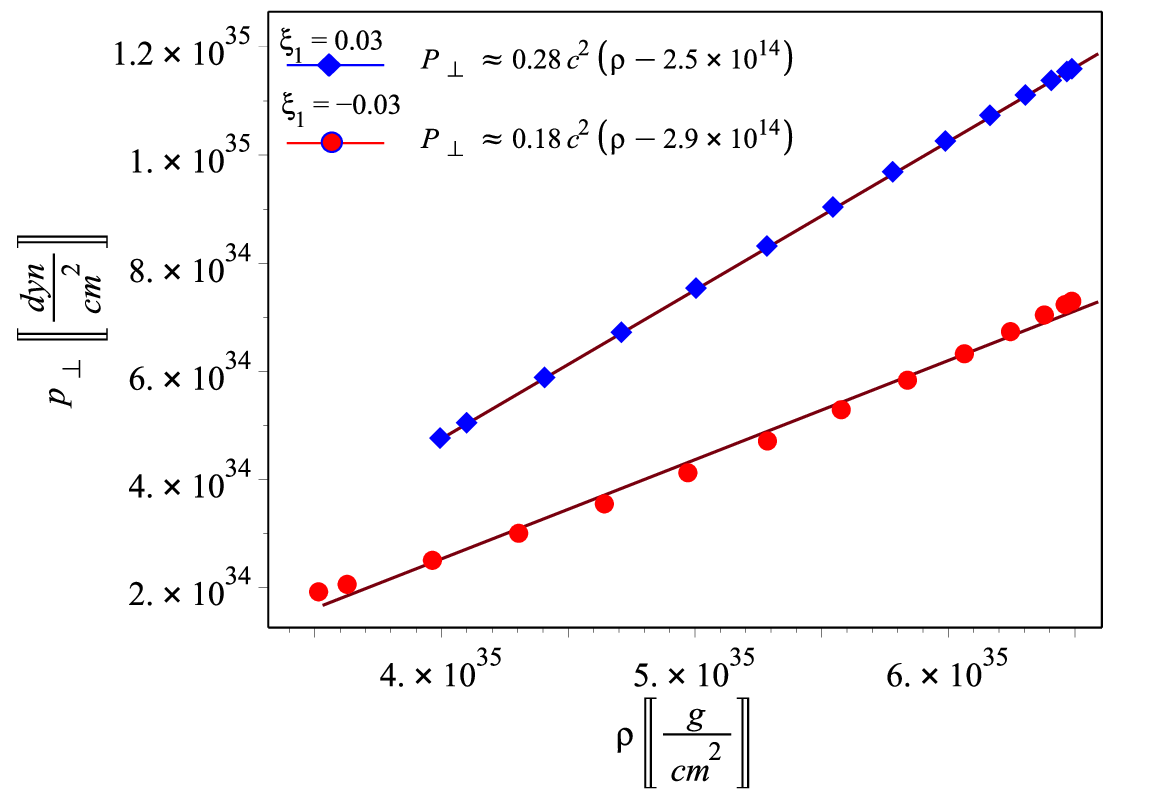}}
\caption{
The set of density and radial pressure data are obtained through the equations presented in Appendix~\ref{Sec:Appendix New B.} when $\xi_1=\pm 0.03$. Similarly, for the tangential form of the EoSs and when $\xi_1=\pm 0.03$ the data show a complete agreement with respect to the linear form of the  model. These linear forms of the EoSs align with those previously derived, specifically those presented in Eq.~\eqref{eq:KB_EoS2}, which satisfies the condition that the relations are appropriate throughout the interior of a pulsar.
}
\label{Fig:EoS}
\end{figure}

The mysterious composition of matters in the cores of a neutron star, where the densities far exceed the levels of the nuclear saturation density unattainable by Earth-based laboratories, has been explored. Although the EoS for matters in a neutron star is not understood so clearly, the astrophysical observations of radius and mass offer the opportunity to put limits on it, or at minimum, rule out certain possibilities.
  Using the numerical data presented in Subsection \ref{Sec:obs_const} for the star ${\mathcal J0740+6620}$ and the equation of state derived from the quadratic form of $f(\mathcal{Q})$ theory, as specified by equations presented in Appendix~\ref{Sec:Appendix New B.}, we generate the series depicted in Fig.~\ref{Fig:EoS}.

The data show an agreement with the linear shape  of the model when, i.e., $\xi_1=\pm 0.03$. When $\xi_1=0.03$ the efficient  equations are written as $P_\perp \text{[dyn/cm$^2$]}\approx 0.28c^2(\epsilon-2.5\times 10^{14}\text{[g/cm$^3$]})$ and $P_r \text{[dyn/cm$^2$]}\approx 0.47 c^2(\epsilon-4.5 \times 10^{14}\text{[g/cm$^3$]})$. When $\xi_1=-0.03$ the efficient  equations are $P_r \text{[dyn/cm$^2$]}\approx 0.25 c^2(\epsilon-4 \times 10^{14}\text{[g/cm$^3$]})$ and $P_\perp \text{[dyn/cm$^2$]}\approx 0.18c^2(\epsilon-2.9\times 10^{14}\text{[g/cm$^3$]})$. These equations are consistent with the EoSs derived by Eq.~\eqref{eq:KB_EoS2} previously, namely, $P_\perp \text{[dyn/cm$^2$]}\approx 0.26c^2(\epsilon-2.3\times 10^{14}\text{[g/cm$^3$]})$ and $P_r \text{[dyn/cm$^2$]}\approx 0.50c^2(\epsilon-4.8 \times 10^{14}\text{[g/cm$^3$]})$ for $\xi_1=0.03$, and $P_\perp \text{[dyn/cm$^2$]}\approx 0.2c^2(\epsilon-2.9\times 10^{14}\text{[g/cm$^3$]})$ and $P_r \text{[dyn/cm$^2$]}\approx 0.29 c^2(\epsilon-3.5 \times 10^{14}\text{[g/cm$^3$]})$ for $\xi_1=-0.03$.
%
\begin{figure*}[t]
\subfigure[~Compactness]{\label{fig:Comp}\includegraphics[scale=0.4]{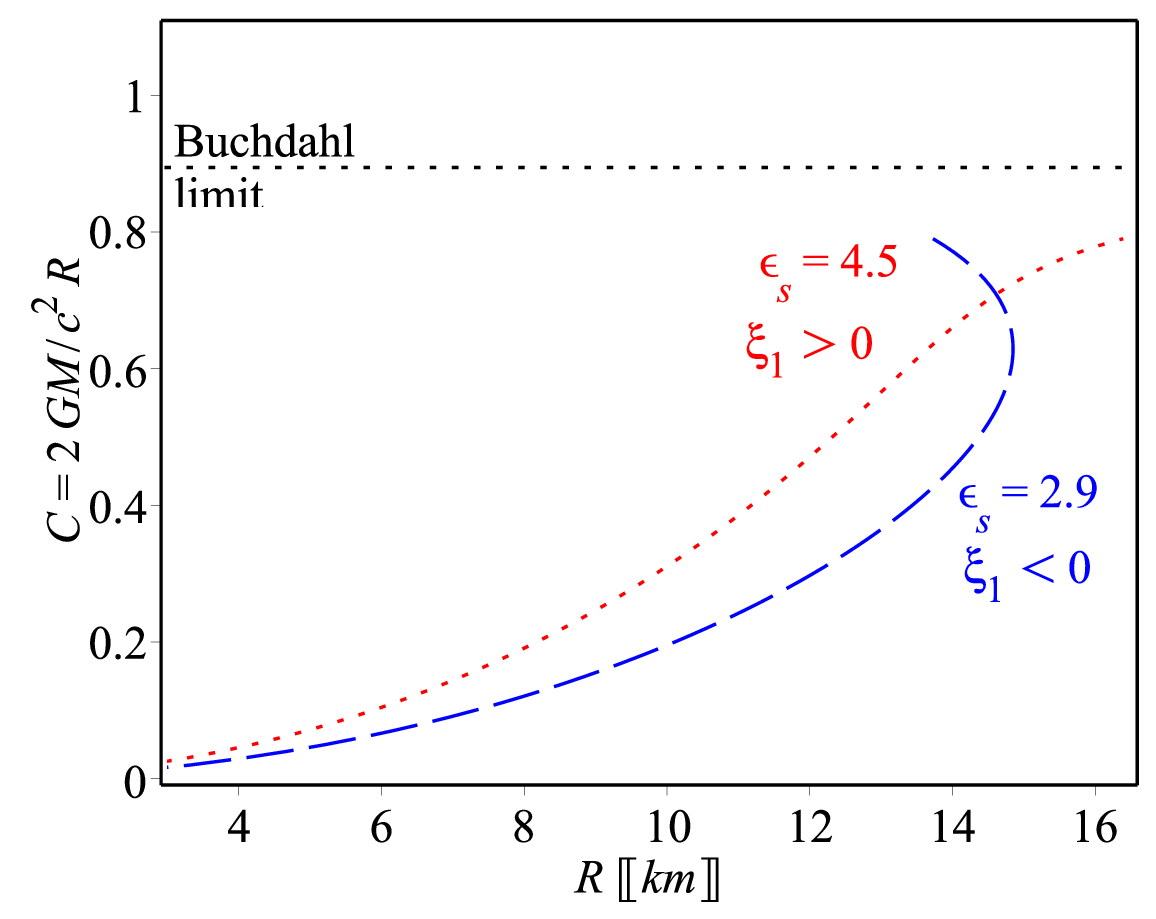}}\hspace{0.5cm}
\subfigure[~Mass-radius relation]{\label{fig:MR}\includegraphics[scale=.4]{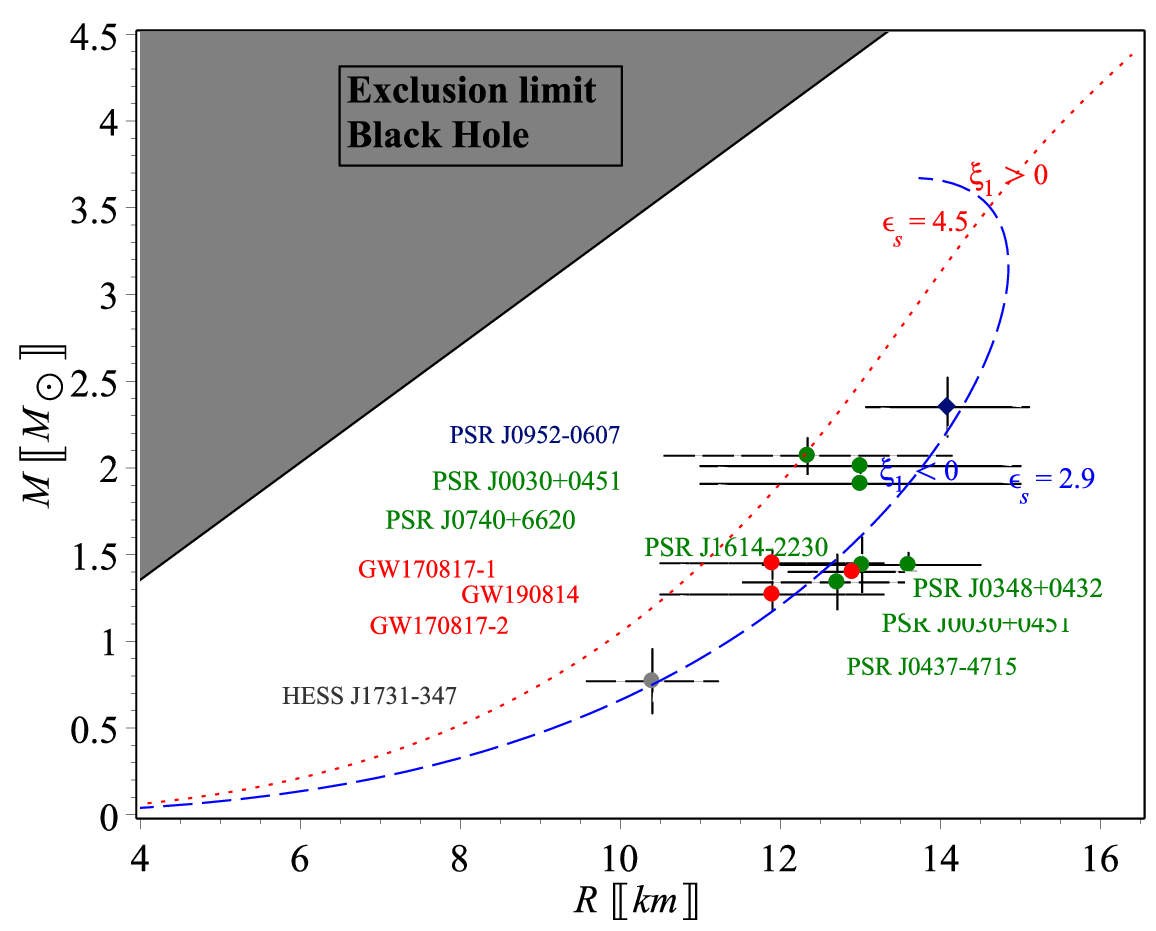}}
\caption{The compactness is shown in Fig.~\subref{fig:Comp}. The dashed horizontal line represents the Buchdahl constraint if the compactness is $C=8/9$. The behavior of the Compactness-radius relation linked to the fits of the EoS is presented in Fig.~\ref{Fig:EoS}. Additionally, the Compactness-radius behavior is demonstrated for the surface densities of $\epsilon_s=2.9\times 10^{14}$ g/cm$^3$ and $\epsilon_s=4.5\times 10^{14}$ g/cm$^3$. Notably, when $\xi_1=\pm0.03$ the high  compactness values remain below Buchdahl upper limit for both densities, highlighting how the quadratic form of $f(\mathcal{Q})$ gravity differs from general relativity. Furthermore, the mass-radius (MR) relation is shown in Fig.~\subref{fig:MR}. Similarly to the case for $\pm \xi_1$, the MR plots do not intersect the limit of Buchdahl.}
\label{Fig:CompMR}
\end{figure*}
%


In this study, concerning the stellar ${\mathcal J0740+662}$, the limit of Buchdahl is explored: when $\xi_1=\pm0.03$ we derive $C\lesssim 0.871$. The data from observation are consistent with ours in the frame of the quadratic form of $f(\mathcal{Q})$ which presents an extra force in the TOV equation in the hydrodynamic equilibrium. It is found that the pulsar can support a larger mass and achieve higher compactness values due  to this extra force, which contradicts the gravitational force. Since the Buchdahl limit is slightly altered in both scenarios,
we employ the usual constraint of $C\leq 8/9$, as shown with the dashed line in Fig. \ref{Fig:CompMR}\subref{fig:Comp}. We suppose that the surface density of $\epsilon_\text{s}=2.7\times 10^{14}$ g/cm$^3$ for both $\xi_1=\pm0.03$ as fixed by the fit of EoS. We calculate the density configuration for the different cases of the compactness parameter when $0 \leq C \leq 1$ utilizing Eq.~({\color{blue} \bf $\mathrm{B}1$}) to obtain the $R$.
 It is interesting to observe that for positive and negative values of $\xi_1$ the highest compactness value still under the Buchdahl upper value, demonstrated in Fig.~\ref{Fig:CompMR}\subref{fig:Comp}.

In Fig.~\ref{Fig:CompMR}\subref{fig:MR}, the mass-radius curves is displayed for both cases of $\xi_1$  based on the previously derived optimal EoS. Within these plots, $M$ is fixed using the junction constraints \eqref{eq:bo}, i.e., $M= \frac{c^2 L_s}{2 G} (1-e^{-s_2})$. Hence, we choose the surface density to be $\epsilon_s=2.5\times 10^{14}$ g/cm$^{3}$ when $\xi_1=0.03$ resulting a highest  mass of approximately greater than $5 M_\odot$ at a radius of $L_s=22$ km.
For a boundary density of $\epsilon_s=2.9\times 10^{14}$ g/cm$^{3}$ and $\xi_1=-0.03$, this configuration results in a maximum mass of approximately greater than $3 M_\odot$ at a radius of approximately $L_s\approx 15$ km. Such values correspond precisely to the observations for the pulsar ${\mathcal J0740+6620}$.

\section{Conclusions}\label{Sec:Conclusion}

Pulsars are crucial tools in astronomy, providing valuable insights for testing and refining physical theories, including general relativity and alternative gravitational models. By exploring the behavior of a pulsar under the influence of exotic terms in the Einstein action, such as the non-metricity term $f(\mathcal{Q})$, we can deepen our understanding of the fundamental physical principles that govern these astronomical objects. By investigating how pulsars respond to modifications in gravity, we can gain important knowledge about the nature of spacetime, matter, and the forces that shape the universe.

In this study, we have explored the higher-order $f(\mathcal{Q})$ gravitational theory, analyzing its implications for astrophysical objects, particularly pulsars, in light of observational data. We considered a more reliable scenario with an anisotropic energy-momentum tensor, which is expected to better describe the dense matter found within stars. Additionally, we introduced an assumption that in the inner regions of a static pulsar with spherical symmetry, the KB form can be satisfied, ensuring the smoothness of the spacetime. Specifically, we have used observational data from the pulsar ${\mathcal PSR J0740+6620}$, with a mass of $M=2.07 \pm 0.11 M_\odot$ and a radius of $L=12.34^{+1.89}_{-1.67}$ km, as obtained from NICER+XMM observations \cite{Legred:2021hdx}, to further constrain our model.

In essence, an additional force is involved by the anisotropy into the TOV equation. The force causes a star to expand and makes a star in equilibrium to achieve greater mass and compactness than in an isotropic scenario. However, for dense, compact stars, the sound speed in the radial direction is lower compared to the isotropic case. This reduction in sound speed results in making the equation of state more flexible and it is compatible with the observed deformability in the signals of the gravitational waves.

Our findings reveal that in the quadratic expression of $f(\mathcal{Q})$, the allowable range is constrained to $|\xi_1| < 0.03$ (or $|\xi| < 3 \mathrm{km}^2$).  Additionally, we demonstrate the viability of the present model by applying various stability conditions to both the geometric and material components. We also identify the emergence of an additional force due to the quadratic term in $f(\mathcal{Q})$. This force acts as a repulsive effect, especially when
$\xi_1$   is positive, as shown in Fig.~\ref{Fig:TOV}\subref{fig:ve}. The pulsar is able to expand while maintaining its mass near the maximum value due to this force, as illustrated in Fig.~\ref{Fig:CompMR}\subref{fig:Comp}. In contrast, when
$\xi_1$  is negative, this force aids in gravitational collapse, similar to general relativity but with increased intensity. This tendency toward gravitational collapse suggests that the positive $\xi_1$ value in quadratic non-metricity gravity may offer advantages over the general relativity scenario.

Furthermore, we have explicitly displayed the optimal fitting EoS by plotting the relationship of mass-radius, as shown in Fig. \ref{Fig:CompMR}\subref{fig:MR}. If $\xi_1=-0.03$ and a surface density of $\epsilon_s=2.9\times 10^{14}$ g/cm$^{3}$, a greater mass of $M=3.67 M_\odot$ accompanied by a radius of $L_s=14.15$ km can be realized. By utilizing the surface density of $\epsilon_s=4.5\times 10^{14}$ g/cm$^{3}$, the mass was found to be higher at $M=4.45 M_\odot$ and the radius at $L_s=16.59$ km when $\xi_2=0.03$. These values precisely match the ones seen in the pulsar ${\mathcal J0740+6620}$. In addition, it is important to mention that the interaction between strong anisotropy and the quadratic shape of $f(\mathcal{Q})$ gravity (in the case of negative $\xi$) prevents gravitational collapsing and reduces the sound speed for a fluid in a star. This fact yields a higher speed of sound in the radial direction of about $v_r^2\approx 0.47c^2$ at the core, which aligns well with making the equations of state softer, as seen by the observations of gravitational waves. By comparing with the case of general relativity or the quadratic gravity theory with a positive value of $\xi_1$, the value exceeds the proposed maximum conformal higher value of the sound speed, $c_s^2=c^2/3$.

\section*{Acknowledgements}
Kazuharu Bamba acknowledges the support by the JSPS KAKENHI Grant Numbers JP21K03547, 24KF0100.

\appendix
\section{Energy density and pressures}\label{Sec:Appendix New A.}
In this Appendix, we show the representations of $\epsilon$, $P_r$ and
$P_\perp$ in Subsection~\ref{Sec:IIIA}. These expressions are given by
\begin{align}\label{Rho1}
&\epsilon=\frac {{e^{-a}}}{4{r}^{2}{c}^{2}{\kappa}} \left\{ 4\,{e^{a}}-4\,{e^{a-b }}+{e^{a}}a'^{2}\xi+{e^{a}}b'^{2}\xi+{e ^{a-2\,b }}\xi\, a'^{2}+4\,{e^{a-b }} a'\xi\,b'-6\,\xi \,{e^{a-2\,b }} b' a'+2\,{e^{a}} a' \xi\,b' + 6\,{e^{a-b }}\xi\, b'^{2}\right.\nonumber\\
&\left.-7\,\xi\,{e^{a- 2\,b }} b'^{2}+4\,\xi\,{e^{a-2\,b }}a''+4\,\xi\,{e^{a-2\,b }}b'' -2\,{e^{a-b }}\xi\, a'^{2}+4\,\xi\,{e^{a}}a''+4\,\xi\,{ e^{a}}b'' -8\,\xi\,{e^{a-b }}a''-8\,\xi\,{e^{a-b }}b'' +4\,{e^{a-b }} b' r \right\} \,,
\end{align}
\begin{align}\label{pr1}
&P_r=-\frac {{e^{-b }}}{4{r}^{2}{\kappa}} \left\{ \xi\,{e ^{-b }} a'^{3}r+2\, b'r -\xi\, a'^{3}r+{ e^{b }} a'^{2}\xi-2\, a'^{2}\xi+ a'b'{ r}^{2}-2 a' r+2r a' \xi{e^{- b }}b'' +4r \xi\,{e^{-b }} a' a'' + b'^{2}r a'\xi-2r a''\xi b'\right.\nonumber\\
&\left. -2\,r a' \xi\,b''-4\,r\xi\, a'a'' -2\, a'' {r}^{2}- a'^{2}{r}^{2}-2\, b' r\xi\,{e^{-b }} a'^{2}-3\, b'^{2}r a'\xi\,{e^{-b }}+2 \,r a'' \xi \,{e^{-b }}b' +{ e^{b }} b'^{2}\xi+2\,{e^{b }} a' \xi\,b' -2\, a' \xi\,'b \right.\nonumber\\
&\left.+\xi\,{e^{ -b }} a'^{2}-\xi\,{e^{-b }} b'^{2} \right\} \,,
\end{align}
\begin{align}\label{pt1}
&P_\perp=-\frac {{e^{-b }} }{4{r}^{2}{\kappa}}  \left\{ \xi\,{e ^{-b }} a'^{3}r+2\, b' r -\xi\, a'^{3}r+{ e^{b }} a'^{2}\xi-2\, a'^{2}\xi+ a'b' { r}^{2}-2\, a'r+2\,r a'\xi\,{e^{- b }}b'' +4\,r \xi\,{e^{-b }} a'a'' +b'^{2}r a'\xi\right.\nonumber\\
&\left.-2\,r a''\xi\,b' -2\,r a'\xi\,b'' -4\,r\xi\, a'a'' -2\, a''{r}^{2}- a'^{2}{r}^{2}-2\,b' r\xi\,{e^{-b }} a'^{2}-3\, b'^{2}r a' \xi\,{e^{-b }}+2 \,r a'' \xi \,{e^{-b }}b' +{ e^{b }}b'^{2}\xi+2\,{e^{b }} a' \xi\,b'\right.\nonumber\\
&\left. -2\, a'\xi\,b' +\xi\,{e^{ -b }} a'^{2}-\xi\,{e^{-b }} b'^{2} \right\}\,.
\end{align}

\section{Expressions of energy density and pressures for the KB metric}\label{Sec:Appendix New B.}
In this Appendix, we denote the representations of $\epsilon$, $P_r$ and
$P_\perp$ using the form of the ansatz of $a(r)$ and $b(r)$ given by Eq. \eqref{eq:KB}. These representations are shown as
\begin{align}\label{Eq:NewB1}
&  \epsilon= \frac{1}{{L_s}^{2}  {e^{{2 \frac {s_2{r}^{2}}{{L_s}^{2}}}}} {r}^{2}{c}^{2}{ \kappa}}\left\{  \left( \left( s_0+s_2 \right)  \left[ \left( s_0+s_2 \right) {r}^{2}+{L_s}^{2} \right] \xi_1+2{L_s}^{2} \right)   {e^{{\frac {2s_2{r}^{2 }}{{L_s}^{2}}}}}- \left( 2\left( s_0+s_2 \right)  \left[  \left( s_0-3s_2 \right) {r}^{2}+2{ L_s}^{2} \right]\xi_1\right.\right. \nonumber\\
&\left.\left.-2s_2{r}^{2}+{L_s}^{2} \right) {e^{{\frac {s_2{r}^{2}}{{L_s}^{2}}}}}+ \left[ \left(  s_0-7s_2 \right) {r}^{2}+2{L_s}^{2} \right]  \left( s_0+s_2 \right)\xi_1 \right\}  \,,
\end{align}
\begin{align}\label{Eq:NewB2}
&P_r=\frac{1}{L_s^{2} {e^{{\frac {2{r}^{2}s_2}{L_s^{2}}}}}{r}^{2}{ \kappa}}\left\{  \left(L_s^{2}- \left[  \left( s_0+s_2 \right) {r}^{2}+2{L_s}^{2} \right]  \left( s_0+s_2 \right) \xi_1 \right)  {e^{{\frac {2{r}^{2}s_2} {L_s^{2}}}}} + \left( 2 \left( s_0+s_2 \right)  \left(  \left( 3s_0-s_2 \right) {r}^{2}+2L_s{}^{2} \right) \xi_1\right. \right.\nonumber\\
&\left.\left.+2{r}^{2}s_0+L_s^{2} \right) {e^{{\frac {{r}^{2}s_2}{L_s^{2}}}}}-\xi_1 \left[  \left( 5s_0-3s_2 \right) {r}^{2}+2L_s^{2} \right]  \left( s_0+s_2 \right)  \right\} \,,
\end{align}
\begin{align}\label{Eq:NewB3}
 &{  P_\perp}=\frac{1}{{L_s}^{4}{\kappa}{e^{{\frac {2{r} ^{2}s_2}{{L_s}^{2}}}}}}\left\{2{ e^{{\frac {{r}^{2}s_2}{{L_s}^{2}}}}}{r}^{2}\xi_1{s_0}^{3}-{L_s}^{2}\xi_1{e^{2{\frac {{r}^{2}s_2}{{L_s }^{2}}}}}{s_0}^{2}-2{L_s}^{2}\xi_1{e^{2{\frac {{ r}^{2}s_2}{{L_s}^{2}}}}}s_0s_2-{L_s}^{2}\xi_1{ e^{2{\frac {{r}^{2}x_2}{{L_s}^{2}}}}}{s_2}^{2}+{e^{{\frac {{r}^{2}s_2}{{L_s}^{2}}}}}{s_0}^ {2}{r}^{2}+6{e^{{\frac {{r}^{2}s_2}{{L_s}^{2}}}}}{L_s}^{2}{s_0}^{2}\xi_1\right.\nonumber\\
&\left.-2{e^{{\frac {{r}^{2}s_2}{{L_s}^{2} }}}}s_0\xi_1{r}^{2}{s_2}^{2}+6{e^{{\frac {{r}^{2}s_2}{{L_s}^{2}}}}}s_0{L_s}^{2}\xi_1s_2-{ e^{{\frac {{r}^{2}s_2}{{L_s}^{2}}}}}s_0{r}^{2}s_2 +2{e^{{\frac {{r}^{2}s_2}{{L_s}^{2}}}}}s_0{L_s}^{2}-{ e^{{\frac {{r}^{2}s_2}{{L_s}^{2}}}}}s_2{L_s}^{2}-2{r}^ {2}\xi_1{s_0}^{3}+4\xi_1{r}^{2}{s_0}^{2} s_2\right.\nonumber\\
&\left.-5{L_s}^{2}{s_0}^{2}\xi_1-4s_0{L_s}^{2}\xi_1s_2+6s_0\xi_1{r}^{2}{s_2}^{2} +{L_s}^{2}\xi_1{s_2}^{2} \right\}
\,.
\end{align}

\section{The KB framework and the corresponding EoS }\label{Sec:App_1.}
The KB ansatz establishes a connection between pressure and density, from that the EoS obtained from equations in \eqref{eq:KB_EoS} are derived. The coefficients are represented as
\begin{align}
&b_1=\frac {  4\,s_0\, \left( 2\xi_1\,s_0-1 \right) {L_s}^{2}{\kappa}^{2}{c}^{2}-12\,{\xi_1}\,{s_2}^{2}{L_s}^{2}{c}^{2}{ \kappa}^{2}+ \left( {L_s}^{2}{c}^{2}{\kappa}^{2}-4\,s_0\,{L_s}^{2}{c}^{2}{\kappa}^{2}\xi_1 \right) s_2  }{5{L_s}^{2}s_2\, \left( 4\xi_1\,s_0+4\xi_1\,s_2-1 \right) {\kappa}^{2}}\,
,
\end{align}
\begin{align}
&b_2=\frac {16\,{s_2}^{3}\xi_1+ \left( 2+32\,\xi_1s_0 \right) {s_2}^{2}+ \left( 16\,\xi_1\,{s_0
}^{2}+2\,s_0 \right) s_2}{5{L_s}^{2}s_2\, \left(4\xi_1\,s_0+4\xi_1\,s_2 -1\right) {\kappa}^{2}}\,,
\end{align}
\begin{align}
&b_3=\frac {   \left( 12\,s_0\,{L_s}^{2}{c}^{2}{\kappa}^{2}\xi_1+2\,{L_s}^{2}{c}^{2}{\kappa}^{2} \right) {s_2}^{2}-4\,{s_2}^{3}\xi_1\,{L_s}^{2}{c}^{2}{
\kappa}^{2}+ \left(16\,\xi_1\,{s_0}^{2}{L_s}^{2}{c}^{2}{\kappa}^{2}-6\,s_0
\,{L_s}^{2}{c}^{2}{\kappa}^{2} \right) s_2+2\,{s_0}^{2}{L_s}^{2}
{c}^{2}{\kappa}^{2} }{5{L_s}^{2}{s_2}^{2} \left(4\xi_1\,s_0+4\xi_1\,s_2-1 \right) {\kappa}^{2}}
\,,
\end{align}
\begin{align}
&b_4=\frac { 8\left( s_0-\xi_1\,{s_0}^{2} \right)-8\,{s_2}^{4}\xi_1- \left( 1+16\,\xi_1\,s_0 \right) {s_2}^{3} {s_2}^{2}-6\,{s_0}^{2}s_2}{5{L_s}^{2}{s_2}^{2} \left(4\xi_1\,s_0+4\xi_1\,s_2  -1\right) {\kappa}^{2}}
\,.
\end{align}
These representations lead to Eq. \eqref{eq:KB_EoS2} for $v_r^2=c_1$, $\epsilon_1=\epsilon_s=-c_2/c_1$, $v_t^2=c_3$ and $\epsilon_2=-c_4/c_3$.

\section{The gradients of density and pressure}\label{Sec:App_2.}
Referring to the density and pressures of matter derived from the quadratic
terms of equations in Appendix~\ref{Sec:Appendix New B.},
the gradients in terms of the radial distance are given by
\begin{align}
&\epsilon'=\frac{1}{{r}{L_s}^{4} {e^{{ \frac {2s_2{r}^{2}}{{L_s}^{2}}}}}{c}^{2}{\kappa}^{ 2}}\left\{  \left( 2 \left( s_0+s_2 \right)  \left( 2{s_2}^{2}{r}^{2}+ \left( 2{r}^{2}s_0+5{L_s}^{2} \right) s_2 +{L_s}^{2}s_0 \right) \xi+4{L_s}^{2}s_2 \right) {e^{{\frac {2s_2{r}^{2}}{{L_s}^{2}}}}} + \left( 4 \left( 3{s_2}^{2}{r}^{2}-\left( {r}^{2}s_0- {L_s}^{2} \right) s_2 \right.\right.\right.\nonumber\\
&\left.\left.\left.-{L_s}^{2}s_0 \right)\left( s_0+s_2 \right) \xi_1+4\,{s_2}^{2}{r}^{2}+2\,{L_s}^{2}s_2 \right) {e^{{\frac {s_2\,{r}^{2}}{{L_s}^{2}}}}}+2\, \left( s_0-7\,s_2 \right) \xi_1\, \left( s_0+s_2 \right) {L_s}^{2} \right\}\, ,
\end{align}
\begin{align}
  & P'_r=\frac{1}{{r}{L_s}^{4}{e^{{\frac {2s_2{r} ^{2}}{{L_s}^{2}}}}}{\kappa}^{2}}\left(  \left( -2 \left( s_0+s_2 \right)  \left( 2{s_2}^{2}{r}^{2}+ \left( 2{r}^{2}s_0+5{L_s}^{2} \right) s_2+{L_s}^{2}s_0 \right) \xi_1-4{L_s}^{2}s_2 \right) {e^{{\frac {2s_2{r}^{2}}{{L_s}^{2}}}}}+ \left( 12 \left( s_0+s_2 \right)  \left({L_s}^{2}s_0 \right.\right.\right.\nonumber\\
&\left.\left.\left.+ \left( {r}^{2}s_0+\frac{1}3{L_s}^{2} \right) s_2 -\frac{1}3{s_2}^{2}{r}^{2} \right) \xi_1+ \left( 2{L_s}^ {2}+4{r}^{2}s_0 \right) s_2+4{L_s}^{2}s_0 \right) { e^{{\frac {s_2{r}^{2}}{{L_s}^{2}}}}}-10 \left( s_0-\frac{3}5s_2 \right) \xi_1 \left( s_0+s_2 \right) { L_s}^{2} \right) \,,
\end{align}
\begin{align}
&P'_\perp=\frac{4r}{{L_s} ^{6} {e^{{\frac {2s_2{r}^{2}}{{L_s}^{2}}}}}{\kappa}^{2}} \left\{\left[ \xi_1 \left( s_2{r}^{2}+{L_s}^{2} \right) {s_0}^{3}+ \left[  \left( 3\xi_1{L_s}^{2}+\frac{1}2 {r}^{2} \right) s_2+\frac{1}2{L_s}^{2} \right] {s_0}^{2}+ \left[\frac{1}2{L_s}^{2}s_2+ \left( 2\xi_1{L_s}^{2} -\frac{1}2{r}^{2}\right) {s_2}^{2}\right.\right.\right.\nonumber\\
&\left.\left.\left.  - {s_2}^{3}\xi_1{r}^{2}\right] s_0-1/2{L_s}^{2}{s_2}^{2} \right] {e^{{\frac {s_2 {r}^{2}}{{L_s}^{2}}}}}- \left( s_0+s_2 \right) s_0 \left( s_0-3s_2 \right) \xi_1{L_s}^{2}  -\xi_1{L_s}^{2}s_2 \left( s_0+s_2 \right) ^{2} {e^{{\frac {2s_2{r}^{2}}{{L_s}^{2}}}}}\right\}
\,.
\end{align}


\end{document}